\documentclass[pre,aps,onecolumn,groupedaddress]{revtex4-1}
\usepackage{amssymb,epsfig,amsmath,bm,subfigure,graphicx,textcomp,url,float,color,cases}
\usepackage[normalem]{ulem}
\usepackage{mathtools}
\usepackage{fixltx2e}
\usepackage{wasysym}

\begin{document}

\title{Height distribution tails in the Kardar-Parisi-Zhang equation with Brownian initial conditions}

\author{Baruch Meerson}
\email{meerson@mail.huji.ac.il}
\affiliation{Racah Institute of Physics, Hebrew University of Jerusalem, Jerusalem 91904, Israel}

\author{Johannes Schmidt}
\email{schmidt@thp.uni-koeln.de}
\affiliation{Institut f\"{u}r Theoretische Physik, Universit\"{a}t zu K\"{o}ln, Z\"{u}lpicher Str. 77, K\"{o}ln D-50937 Germany}

\begin{abstract}
For stationary interface growth, governed by the Kardar-Parisi-Zhang (KPZ) equation in $1 + 1$ dimensions, typical fluctuations of the interface height at long times are described by the Baik-Rains distribution. Recently Chhita \textit{et al}. \cite{CFS} used
the totally asymmetric simple exclusion process  (TASEP) to study the height fluctuations
in systems of the KPZ universality class for Brownian interfaces with arbitrary diffusion constant. They showed that there is a one-parameter family of long-time distributions, parameterized by the diffusion constant of the initial random height profile. They also computed these distributions numerically by using Monte Carlo (MC) simulations. Here we address this problem analytically and focus on the distribution tails at short times. We
determine the (stretched exponential) tails of the height distribution by applying the Optimal Fluctuation Method (OFM) to the KPZ equation. We argue
that, by analogy with other initial conditions,  the ``slow" tail holds at arbitrary times and therefore provides a proper asymptotic to the family of long-time distributions studied in Ref. \cite{CFS}. We verify this hypothesis by performing large-scale MC simulations of a TASEP with a parallel-update rule. The ``fast" tail, predicted by the OFM, is also expected to hold at arbitrary times,
at sufficiently large heights.

\end{abstract}

\maketitle
\tableofcontents

%#$#$#$#$#$#$#$#$#$#$#$#$#$#$#$#$#$#$#$#$#$#$#$#$#$#$#$#$#$#$#$#$#$#$#$#$#$#$#$#$#$#$#$#$#$#$#$#$#$#$#$#$#$#$#$#$#$#$#$#$#$#$#$#$#$#$#$
%#$#$#$#$#$#$#$#$#$#$#$#$#$#$#$#$#$#$#$#$#$#$#$#$#$#$#$#$#$#$#$#$#$#$#$#$#$#$#$#$#$#$#$#$#$#$#$#$#$#$#$#$#$#$#$#$#$#$#$#$#$#$#$#$#$#$#$
%#$#$#$#$#$#$#$#$#$#$#$#$#$#$#$#$#$#$#$#$#$#$#$#$#$#$#$#$#$#$#$#$#$#$#$#$#$#$#$#$#$#$#$#$#$#$#$#$#$#$#$#$#$#$#$#$#$#$#$#$#$#$#$#$#$#$#$
\section{Introduction}
\label{sec:Intro}

The Kardar-Parisi-Zhang (KPZ) equation \cite{KPZ}  describes an important universality class
of non-equilibrium surface growth \cite{JS:KRUG_Book_GROWTH,JS:BARABASI_GROWTH,HHZ,Krug,Corwin,QS,HHT,S2016}. In $1+1$ dimension the KPZ equation,
\begin{equation}\label{eq:KPZoriginal}
\partial_{t}h=\nu \partial^2_{x}h+\frac{\lambda}{2}\left(\partial_{x}h\right)^2+\sqrt{D}\,\xi(x,t),
\end{equation}
governs the evolution of the interface height  $h(x,t)$. The system is driven by a Gaussian white noise  $\xi(x,t)$ with zero mean
and covariance
\begin{equation}\label{eq:LD040}
\left\langle\xi(x,t)\xi(x^\prime,t^\prime)\right\rangle=\delta\left(x-x^\prime\right)\, \delta\left(t-t^\prime\right)\,.
\end{equation}
At long times the characteristic interface width grows as $t^{1/3}$, whereas the correlation length in the $x$-direction grows as $t^{2/3}$, in agreement with experiments \cite{experiment1}. In recent years the focus of interest in the $1+1$ KPZ equation  shifted toward more detailed quantities. One of them is the complete one-point probability distribution  ${\mathcal P}_t(H)$ of the height $H$ at time $t$ at a given point $x$ \cite{Corwin,QS,HHT,S2016,shifted}.  Several groups obtained exact representations, in term of Fredholm determinants,  for a generating function of  ${\mathcal P}_t(H)$
at arbitrary $t>0$. These important results were derived for several types of initial conditions, and for some of their combinations.
Three basic initial conditions are infinitely narrow wedge (which quickly becomes a ``droplet") \cite{SS,CDR,Dotsenko,ACQ,Corwin}, flat interface \cite{CLD}, and stationary interface \cite{IS,Borodinetal}. At long times, and for typical fluctuations, ${\mathcal P}_t(H)$ converges in these cases to the Tracy-Widom (TW) distribution for the Gaussian unitary ensemble (GUE)  \cite{TW} for the droplet, to the TW distribution for the Gaussian orthogonal ensemble (GOE)   \cite{TWGOE} for flat interface, and to the Baik-Rains (BR) distribution \cite{BR} for stationary interface. Ingenious experiments with liquid-crystal turbulent fronts confirmed these theoretical predictions \cite{experiment2}.

The unexpectedly high sensitivity of the long-time behavior of ${\mathcal P}_t(H)$ to initial conditions prompted interest in the \emph{domains of attraction} of the three basic (GUE TW, GOE TW and BR) long-time distributions, and in the possibility of additional asymptotic distributions \cite{KMS,QR,CFS}. In particular, Chhita, Ferrari and Spohn \cite{CFS} studied the case when the interface at $t=0$ is described by a Brownian motion  with arbitrary
diffusion constant. For one particular value of this diffusion constant these initial data
correspond to the stationary case, whereas a vanishing diffusion constant corresponds to flat interface \cite{CFS}.  The authors \cite{CFS} showed, via analytical arguments and MC simulations of the totally asymmetric exclusion process (TASEP), that the typical height fluctuations at long times (correspondingly, fluctuations of integrated current in the TASEP) are described
by a new one-parameter family of distributions, parameterized by the diffusion constant of the Brownian interface at $t=0$.
Except for the two special values of the diffusion constant, mentioned above, this family of distributions is known only numerically \cite{CFS}.

The last two years have witnessed a growing interest in \emph{large deviations} of the KPZ interface height, described by distant tails  of ${\mathcal P}_t(H)$ \cite{MKV,KMS,JKM,DMRS,DMS,SMP,KD}. Even when exact representations for ${\mathcal P}_t(H)$ are available, extracting the tails from them is technically challenging \cite{DMRS,DMS,SMP,KD}.
A viable alternative is provided by the Optimal Fluctuation Method (OFM). Before the advent of exact representations, Kolokolov and Korshunov \cite{KK2007,KK2008,KK2009}  used the OFM to evaluate  the tails of ${\mathcal P}_t(H)$ for the flat initial condition.  The OFM is a variant of WKB approximation (after Wentzel, Kramers and Brillouin) that goes back to Refs. \cite{Halperin,Langer,Lifshitz,LGP} in condensed matter physics. It is also known as the instanton method in turbulence \cite{turb1,turb2,turb3}, the Macroscopic Fluctuation Theory in lattice gases \cite{MFTreview},  and WKB methods in reaction-diffusion systems \cite{EK,MS2011}. In the context of the KPZ and Burgers equations the OFM is also known as the Weak Noise Theory \cite{Fogedby1998,Fogedby1999,Fogedby2009}.

The crux of the OFM is the ``optimal fluctuation":   the most likely history of the interface height $h(x,t)$ and  the most likely realization of the noise $\xi(x,t)$ which dominate the contribution to ${\mathcal P}_t(H)$. The OFM involves a saddle-point evaluation of the path integral for the KPZ equation
(\ref{eq:KPZoriginal}), constrained by the specified large deviation and the initial and boundary conditions. A necessary (but not always sufficient) condition for the OFM validity demands that the
``classical action", resulting from the saddle-point calculation, be sufficiently large. By now all three basic initial conditions (and the parabolic initial condition which interpolates between the flat and the droplet cases) have been studied with the OFM \cite{KK2007,KK2009,MKV,KMS,JKM}. One general outcome of these studies (see also Refs. \cite{DMRS,DMS,KD} where exact representations
were used instead) concerns the slow tail ($\lambda H>0$) of ${\mathcal P}_t(H)$.
This tail behaves as a stretched exponential,
\begin{equation}\label{eq:negativetail}
-\ln \mathcal{P}_t (H)\simeq \frac{f\,\nu |H|^{3/2}}{D |\lambda|^{1/2} t^{1/2}},
\end{equation}
where $f=f_0=8\sqrt{2}/3$ for the droplet and flat initial conditions (and, more generally, for a broad class of deterministic initial conditions \cite{KMS}), and $f=f_1=4\sqrt{2}/3$ for stationary initial condition.
Importantly, these asymptotics correctly describe the corresponding tails of the typical fluctuations at long times, governed by the GUE and GOE TW distributions, and by the BR distribution, respectively.

In this work we employ the OFM to determine analytically the  $\lambda H>0$ tails of the family of height distributions  ${\mathcal P}_t(H,\sigma)$ observed in Ref. \cite{CFS} for general Brownian initial interfaces. Here the parameter $\sigma$ is the square root of the ratio of the diffusion constant of the Brownian interface and the unique diffusion constant that describes
stationary initial condition. As we show, the  $\lambda H>0$ tails have the following form:
\begin{equation}\label{eq:negativetailsigma}
-\ln \mathcal{P}_t (H,\sigma)\simeq \frac{f(\sigma)\,\nu |H|^{3/2}}{D |\lambda|^{1/2} t^{1/2}} .
\end{equation}
The function $f(\sigma)$ is given by Eq.~(\ref{eq:fsigma}) below, and its graph is shown in Fig.~\ref{fig:f}. As to be expected, $f(\sigma)$ satisfies the relations $f(0)=f_0$ and $f(1)=f_1$, where $f_0$ and $f_1$ were introduced immediately after Eq.~(\ref{eq:negativetail}). By analogy with all previously studied initial conditions, we argue that Eqs.~(\ref{eq:negativetailsigma}) and ~(\ref{eq:fsigma}) hold at all times $t>0$. Therefore, they should correctly describe the  $\lambda H>0$  tails of the family of distributions studied in Ref. \cite{CFS}.
To test this prediction we performed  a large scale MC simulations of the TASEP. In order to resolve the slow
distribution tail for different $\sigma$, we modified the simulation algorithm of Ref. \cite{CFS} by using a parallel,
rather than a random sequential, update rule.
Our simulation results are in very good agreement with Eqs.~(\ref{eq:negativetailsigma}) and ~(\ref{eq:fsigma}), as can be seen in Fig.~\ref{fig:JS:f_sigma_fit_small_sigma} below.

We also briefly address  the  $\lambda H<0$ large-deviation tail of the height distributions  ${\mathcal P}_t(H,\sigma)$.  In the leading order, this tail is independent of $\nu$ and $\sigma$ and given by Eq.~(\ref{eq:linearcond}) below. As time increases, this $H^{5/2}/t^{1/2}$ tail is ``pushed", by an expanding $H^3/t$ tail, toward progressively larger heights.

Here is how we structured the rest of the paper. In Sec. \ref{sec:scaling} we outline the OFM formalism for the KPZ equation. In Sec. \ref{sec:small-H} we briefly describe the short-time behavior
of \emph{typical} height fluctuations, where the KPZ nonlinearity is small. Section \ref{sec:negative H} addresses the $\lambda H>0$ tail of ${\mathcal P}_t(H,\sigma)$ and presents
a derivation of Eqs.~(\ref{eq:negativetailsigma}) and (\ref{eq:fsigma}) -- the main analytic result of this work. In Sec.~\ref{sec:positive H} we obtain the \emph{large-deviation} $\lambda H<0$ tail of  ${\mathcal P}_t(H,\sigma)$. In Sec.~\ref{sec:longtime} we discuss the behavior of the whole distribution  ${\mathcal P}_t(H,\sigma)$ at long times. In Sec.~\ref{sec:TASEP} we present the results of our TASEP simulations and compare them with Eqs.~(\ref{eq:negativetailsigma}) and (\ref{eq:fsigma}).  Section \ref{sec:discussion}
presents a brief summary of this work. A more detailed description of our MC simulations is presented in two Appendices.

%#$#$#$#$#$#$#$#$#$#$#$#$#$#$#$#$#$#$#$#$#$#$#$#$#$#$#$#$#$#$#$#$#$#$#$#$#$#$#$#$#$#$#$#$#$#$#$#$#$#$#$#$#$#$#$#$#$#$#$#$#$#$#$#$#$#$#$
%#$#$#$#$#$#$#$#$#$#$#$#$#$#$#$#$#$#$#$#$#$#$#$#$#$#$#$#$#$#$#$#$#$#$#$#$#$#$#$#$#$#$#$#$#$#$#$#$#$#$#$#$#$#$#$#$#$#$#$#$#$#$#$#$#$#$#$
%#$#$#$#$#$#$#$#$#$#$#$#$#$#$#$#$#$#$#$#$#$#$#$#$#$#$#$#$#$#$#$#$#$#$#$#$#$#$#$#$#$#$#$#$#$#$#$#$#$#$#$#$#$#$#$#$#$#$#$#$#$#$#$#$#$#$#$
\section{Optimal Fluctuation Method}
\label{sec:scaling}

Let  $T$ be the observation time of a specified interface height $h(x=0,t=T)=H$ \cite{shifted}. Rescaling time, $t/T\to t$, the coordinate, $x/\sqrt{\nu T} \to x$, and the interface height, $|\lambda| h/\nu\to h$, we can rewrite Eq.~(\ref{eq:KPZoriginal}) as
\begin{equation}\label{eq:KPZrescaled}
\partial_{t}h=\partial^2_{x}h-\frac{1}{2} \left(\partial_{x}h\right)^2+\sqrt{\epsilon} \,\xi(x,t),
\end{equation}
where $\epsilon=D\lambda^2 T^{1/2}/\nu^{5/2}$ is the rescaled noise magnitude \cite{MKV}, and we have assumed without loss of generality that $\lambda<0$ \cite{signlambda}. From now on, we will use the rescaled variables and the $\lambda<0$ convention unless stated otherwise.

We want to evaluate the probability distribution  $\mathcal{P}_T(H,\sigma)$ of observing $h(0,1)=H$,
under the condition that the initial interface $h(x,0)$ is a random realization of two-sided Brownian motion with an arbitrary rescaled diffusion constant $0<\sigma^2<\infty$:
\begin{equation}\label{arbitrarysigma}
h(x,t=0) = \sigma B(x),
\end{equation}
where $B(x)$ is the two-sided Wiener process with diffusion constant $1$, pinned at $x=0$.
The particular case $\sigma=1$ corresponds to the stationary interface of Eq.~(\ref{eq:KPZrescaled}).
For $\epsilon \ll 1$, that is at short times, the whole
distribution  $\mathcal{P}_T(H,\sigma)$ can be determined  via a saddle-point evaluation of a path integral for Eq.~(\ref{eq:KPZrescaled}). This procedure brings a minimization problem for the action functional \cite{Fogedby1998,Fogedby1999,Fogedby2009,KK2007,MKV,KMS,JKM}. It is crucial that, sufficiently far in the tails of $\mathcal{P}_T(H,\sigma)$,  $\epsilon$ does not have to be small,
and the saddle-point approximation holds at long times. This is the essence of the OFM for the KPZ equation. Specifics of the random (Brownian) initial conditions is in that one has to minimize the \emph{joint} action functional $s=s_{\text{dyn}}+s_{\text{in}}$, where
\begin{eqnarray}\label{eq:actn}
s_{\text{dyn}}&=& \frac{1}{2}\int_{0}^{1}dt\int_{-\infty}^{\infty} dx \left[\partial_{t} h-\partial_{x}^2 h+\frac{1}{2} \left(\partial_{x} h\right)^2\right]^2
\end{eqnarray}
is the dynamical contribution to the action, and
\begin{equation}\label{eq:s0}
s_{\text{in}} = \frac{1}{\sigma^2} \int_{-\infty}^{\infty}  dx \,(\partial_x h)^2|_{t=0}
\end{equation}
is the  ``fluctuation cost" of  initial height profiles \cite{JKM,DG2009}.  The variational procedure closely follows that of Ref. \cite{JKM}, where the particular case of $\sigma=1$ was considered. The ensuing Euler-Lagrange equation can be recast into Hamilton's equations
for two  canonically conjugated fields: the optimal interface profile $h(x, t)$ and the optimal realization of the noise $\xi(x,t)$, which we call $p(x,t)$:
\begin{eqnarray}
% \nonumber to remove numbering (before each equation)
  \partial_{t} h &=& \frac{\delta \mathcal{H}}{\delta p} = \partial_{x}^2 h -\frac{1}{2} \left(\partial_x h\right)^2+p ,  \label{eq:h}\\
  \partial_{t}p &=& - \frac{\delta \mathcal{H}}{\delta h} = - \partial_{x}^2 p - \partial_x \left(p \partial_x h\right) ,\label{eq:Leqrho}
\end{eqnarray}
where
\begin{equation}\label{eq:hamiltonian}
\mathcal{H} = \int dx p\left[\partial_x^2 h-\frac{1}{2} \left(\partial_x h\right)^2+\frac{p}{2}\right]
\end{equation}
is the Hamiltonian.
The condition
\begin{equation}\label{eq:conditionH}
h(x=0,t=1)=H
\end{equation}
leads to a non-trivial boundary condition for $p(x,t=1)$:
\cite{KK2007,MKV}
\begin{equation}\label{eq:pT}
    p(x,t=1)=\Lambda \,\delta(x),
\end{equation}
where the temporary parameter $\Lambda$ should be ultimately expressed via $H$ and $\sigma$.

The boundary condition at $t=0$ follows from the variation of the joint action functional $s$ over $h(x,t=0)$, and takes
the form
\begin{equation}
 p(x,t=0)+ \frac{2} {\sigma^2} \partial_x^2 h(x,t=0) = \Lambda \delta(x), \label{eq:BC0}
\end{equation}
where the first term on the l.h.s. comes from the variation of
$s_{\text{dyn}}$ (via integration by parts in time of the term $\partial_t h$), whereas the second term comes from the variation of $s_{\text{in}}$.
The boundary condition (\ref{eq:BC0}) generalizes the boundary condition for $\sigma=1$ obtained in Ref. \cite{JKM}. A nonzero probability demands that $p(x,t)$ and $\partial_x h(x,0)$ vanish sufficiently rapidly at $|x|\to \infty$. Finally, the initial interface is pinned at the origin:
\begin{equation}\label{eq:pinned}
h(x=0,t=0)=0.
\end{equation}
Once the optimal path is determined, we can evaluate $s=s_{\text{dyn}}+s_{\text{in}}$, where $s_{\text{dyn}}$ can be rewritten as
\begin{equation}
% \nonumber to remove numbering (before each equation)
s_{\text{dyn}} =\frac{1}{2}\int_0^1 dt \int_{-\infty}^{\infty}   dx\, p^2 (x,t). \label{eq:action1}
\end{equation}
This determines $\mathcal{P}_T(H,\sigma)$ up to a pre-exponential factor: $-\ln \mathcal{P}_T(H,\sigma)\simeq s(H,\sigma)/\epsilon$. In the dimensional variables
\begin{equation}\label{eq:actiondgen}
-\ln \mathcal{P}_T(H,\sigma)\simeq \frac{\nu^{5/2}}{D\lambda^2\sqrt{T}}\,\,
s\left(\frac{|\lambda| H}{\nu},\sigma\right).
\end{equation}
As we already mentioned, at $T\to 0$ the action $s(H,\sigma)$ yields the complete large deviation function of $\mathcal{P}_T(H,\sigma)$. But even at large $T$, sufficiently far tails of the distribution are still described correctly.

%#$#$#$#$#$#$#$#$#$#$#$#$#$#$#$#$#$#$#$#$#$#$#$#$#$#$#$#$#$#$#$#$#$#$#$#$#$#$#$#$#$#$#$#$#$#$#$#$#$#$#$#$#$#$#$#$#$#$#$#$#$#$#$#$#$#$#$
%#$#$#$#$#$#$#$#$#$#$#$#$#$#$#$#$#$#$#$#$#$#$#$#$#$#$#$#$#$#$#$#$#$#$#$#$#$#$#$#$#$#$#$#$#$#$#$#$#$#$#$#$#$#$#$#$#$#$#$#$#$#$#$#$#$#$#$
%#$#$#$#$#$#$#$#$#$#$#$#$#$#$#$#$#$#$#$#$#$#$#$#$#$#$#$#$#$#$#$#$#$#$#$#$#$#$#$#$#$#$#$#$#$#$#$#$#$#$#$#$#$#$#$#$#$#$#$#$#$#$#$#$#$#$#$
\section{Edwards-Wilkinson regime}
\label{sec:small-H}

At small $H$ the optimal path can be determined via a regular
perturbation expansion in the powers of $H$, or $\Lambda$ \cite{MKV,KrMe}. The perturbation series is
$h(x,t)= \Lambda h_1(x,t)+\Lambda^2 h_2(x,t) +\dots$  and $p(x,t)=\Lambda p_1(x,t)+\Lambda^2 p_2(x,t) +\dots$.
This brings an iterative set of coupled \emph{linear} partial differential equations for  $h_i$ and $p_i$ that can be solved order by order \cite{MKV}. The leading order of this perturbation expansion describes the Edwards-Wilkinson region of the $\mathcal{P}_t(H,\sigma)$, where the KPZ nonlinearity can be neglected:
\begin{eqnarray}
% \nonumber to remove numbering (before each equation)
  \partial_t h_1 &=& \partial_x^2 h_1 +p_1,\label{eq:heqEW0}\\
  \partial_t p_1 &=& -\partial_x^2 p_1, \label{eq:peqEW0}
\end{eqnarray}
with the boundary conditions
\begin{equation}\label{eq:linearcond1}
p_1(x,0)+\frac{2} {\sigma^2} \partial_x^2 h_1(x,0) = p_1(x,1) = \delta(x) \quad \text{and}\quad h_1(0,0)=0.
\end{equation}
After a standard algebra one can find  $p_1$ and $h_1$ in terms of $\Lambda$ and $\sigma$ and compute the joint action $s$. Using the condition
$h_1(0,1)=H$, one finally obtains
\begin{equation}\label{eq:linear}
s(H,\sigma) = \frac{\sqrt{\pi } H^2}{\sqrt{2}+\left(2-\sqrt{2}\right) \sigma ^2}.
\end{equation}
Combined with Eq.~(\ref{eq:actiondgen}), this result shows that the body of the short-time height distribution is a Gaussian with a $\sigma$-dependent variance
\begin{equation}
\label{eq:variance}
    \langle H^2 \rangle= \frac{D t^{1/2} \left[\sqrt{2}+(2-\sqrt{2}) \,\sigma^2\right]}{2\sqrt{\pi} \,\nu^{1/2}}
\end{equation}
which describes the Edwards-Wilkinson scaling $\langle H^2 \rangle^{1/2} \sim t^{1/4}$.  For $\sigma \to 0$ and $\sigma=1$ we obtain $s=\sqrt{\pi/2}\, H^2$ and $s=\sqrt{\pi}\, H^2/2$, respectively. These are well-known results for the Edwards-Wilkinson regime of flat and stationary interface, respectively \cite{Krugetal}.  At $\sigma \ll 1$ the cost of creating a non-flat interface is very high, so the optimal initial interface is almost flat. Here the joint action $s$ is dominated by the dynamic
contribution $s_{\text{dyn}}$. At $\sigma\to \infty$ the action $s$ goes to zero, and the variance of $H$ grows indefinitely. This is to be expected, as for large $\sigma$ the fluctuation cost $s_{\text{in}}$ of creating any interface at $t=0$ becomes low. As a result, at $\sigma \to \infty$, the fluctuation cost $s_{\text{in}}$ dominates the joint action $s$.
Fluctuations are only needed in this case to create the optimal initial profile. The following evolution
of $h(x,t)$ proceeds almost deterministically.  As we will see, qualitatively similar features, at $\sigma\ll 1$ and $\sigma\gg 1$, are also observed in the $\lambda H>0$ distribution tail.

In the Edwards-Wilkinson regime the optimal paths $h(x,t)$ and $p(x,t)$ (not shown here) are, at all times,  symmetric functions of $x$.
Higher orders of the perturbation expansion give higher cumulants of the distribution, as was shown in Ref. \cite{MKV}
for the flat initial condition.   Although the KPZ nonlinearity manifests itself already in the second order of the perturbation expansion, the reflection symmetry $x\leftrightarrow -x$  of the optimal paths
is protected in \emph{all} orders of the perturbation theory. Therefore,  within its convergence radius, the perturbation series for $s(H,\sigma)$ comes from a unique solution for the optimal path which obeys the reflection symmetry.

%#$#$#$#$#$#$#$#$#$#$#$#$#$#$#$#$#$#$#$#$#$#$#$#$#$#$#$#$#$#$#$#$#$#$#$#$#$#$#$#$#$#$#$#$#$#$#$#$#$#$#$#$#$#$#$#$#$#$#$#$#$#$#$#$#$#$#$
%#$#$#$#$#$#$#$#$#$#$#$#$#$#$#$#$#$#$#$#$#$#$#$#$#$#$#$#$#$#$#$#$#$#$#$#$#$#$#$#$#$#$#$#$#$#$#$#$#$#$#$#$#$#$#$#$#$#$#$#$#$#$#$#$#$#$#$
%#$#$#$#$#$#$#$#$#$#$#$#$#$#$#$#$#$#$#$#$#$#$#$#$#$#$#$#$#$#$#$#$#$#$#$#$#$#$#$#$#$#$#$#$#$#$#$#$#$#$#$#$#$#$#$#$#$#$#$#$#$#$#$#$#$#$#$
\section{Distribution tails}
\label{sec:tails}

%&&&&&&&&&&&&&&&&&&&&&&&&&&&&&&&&&&&&&&&&&&&&&&&&&&&&&&&&&&&&&&&&&&&&&&&&&&&&&&&&&&&&&&&&&&&&&&&&&&&&&&&&&&&&&&&&&&&&&&&&&&&&&&&&&&&&&&
%&&&&&&&&&&&&&&&&&&&&&&&&&&&&&&&&&&&&&&&&&&&&&&&&&&&&&&&&&&&&&&&&&&&&&&&&&&&&&&&&&&&&&&&&&&&&&&&&&&&&&&&&&&&&&&&&&&&&&&&&&&&&&&&&&&&&&&
\subsection{$\lambda H>0$ tail}
\label{sec:negative H}

For stationary growth, $\sigma=1$, the OFM solution ceases to be unique
at a finite $H<0$ (to remind the reader, we set $\lambda<0$ in this paper) \cite{JKM}. At a critical value $H=H_c\simeq -3.7$ two additional solutions with a broken reflection symmetry appear. Each of them is a mirror reflection
of the other, and they have a lesser action than the (still existing) symmetric solution. As a result, the short-time large deviation function of the hight
$s(H,1)$ exhibits a singularity at $H=H_c$ which has the character of a (mean-field) second-order phase transition. Very recently, the second-order transition at $H=H_c$, at short times, has been reproduced from an exact representation of the height statistics \cite{KD}. For $-H\gg 1$ the symmetry-broken solutions for $h(x,t)$ and $p(x,t)$ and the resulting action were found analytically \cite{JKM}. A crucial building block
of these asymptotic solutions is a traveling pair of a localized pulse of $p$  (which we call soliton) and a fluctuation-driven (that is, $p$-driven) shock of the interface slope
\begin{equation}
u(x,t)=\partial_x h(x,t).
\end{equation}
At $\sigma\neq 1$ one should expect a \emph{critical line} $H=H_c(\sigma)$ of the phase transition to appear. We relegate the determination of the critical line for future work. Here we focus on the
$-H\gg 1$ tail of the height distribution $\mathcal{P}_T(H,\sigma)$ at arbitrary $\sigma$. For sufficiently large negative $\Lambda$, numerical solutions \cite{numerics} show that traveling soliton-shock solutions continue to play a crucial role here, although the complete solution is more complicated than in the case of $\sigma=1$.
To present the traveling soliton-shock solutions, let us first differentiate Eq.~(\ref{eq:h}) with respect to $x$ and rewrite the OF equations as coupled continuity equations for $u$ and $p$:
\begin{eqnarray}
% \nonumber to remove numbering (before each equation)
  \partial_t u+\partial_x\left(\frac{u^2}{2}-\partial_x u-p\right)&=&0, \label{eq:V}\\
  \partial_{t}p  + \partial_x \left(p u +\partial_xp\right)&=& 0  .\label{eq:rhoV}
\end{eqnarray}
One of the two traveling soliton-shock solutions is of the form $p=p(x-ct+C)$ and $u=u(x-ct+C)$, where
\begin{eqnarray}
% \nonumber to remove numbering (before each equation)
p(\xi)&=& -\frac{\Delta u^2}{4} \, \text{sech}^2 \left( \frac{
   \Delta u \,\xi}{4} \right) \,, \label{eq:rhosol}\\
u(\xi) &=& u_{+}-\frac{\Delta u}{1+e^{\frac{\Delta u \xi}{2}}} \label{eq:Vsol} \,, \\
c &=& \frac{u_{-}+u_{+}}{2}, \label{eq:c}
\end{eqnarray}
$C$ is an arbitrary constant, $\Delta u \equiv u_{+}-u_{-}$, and $u_{-}$ and $u_{+}$ are the asymptotic values of $u(\xi)$ at $-\infty$ and $+\infty$, respectively.  This soliton-shock pair travels to the right with velocity $c$. Importantly,  the solution (\ref{eq:rhosol})  and (\ref{eq:Vsol}) belongs to the invariant manifold
\begin{equation}\label{eq:invariant}
p(x,t) + 2 \partial_x^2 h(x,t) \equiv  p(x,t) + 2 \partial_x u(x,t) = 0
\end{equation}
of Eqs.~(\ref{eq:h}) and (\ref{eq:Leqrho}).

The mirror-symmetric soliton-shock solution travels to the left; it can be obtained by replacing $u_{-}$ by $-u_{-}$ and $u_{+}$ by $-u_{+}$. As particular solutions of Eqs.~(\ref{eq:V}) and (\ref{eq:rhoV}), Eqs.~(\ref{eq:rhosol}) - (\ref{eq:c}) have been known for some time \cite{Fogedby1998,Fogedby1999}. Here we will expose their role in the
problem of height statistics specified by the boundary conditions in space and time. We will also show how these solutions should be amended by additional \emph{fluctuationless} solutions, and how the constants  $u_{-}$, $u_{+}$ and $C$ are determined by $\Lambda$ and $\sigma$, and ultimately by $H$ and $\sigma$. For concreteness, we will do it for the solution traveling to the right.

\begin{figure}
\includegraphics[width=0.45\textwidth,clip=]{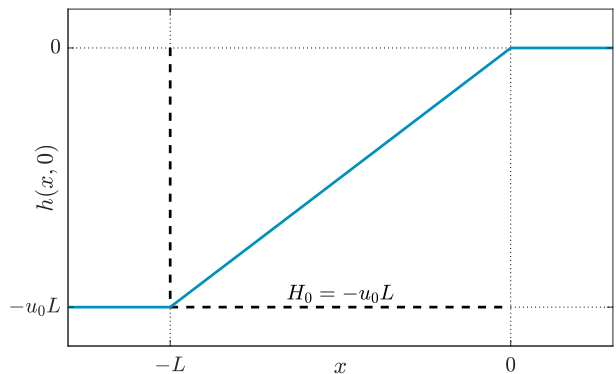}
\caption{The optimal initial interface profile at large positive $\lambda H$. A narrow negative pulse of $p$ (not shown) is located at $x=-L$.}
\label{fig:initial}
\end{figure}

\begin{figure}
\includegraphics[width=0.45\textwidth,clip=]{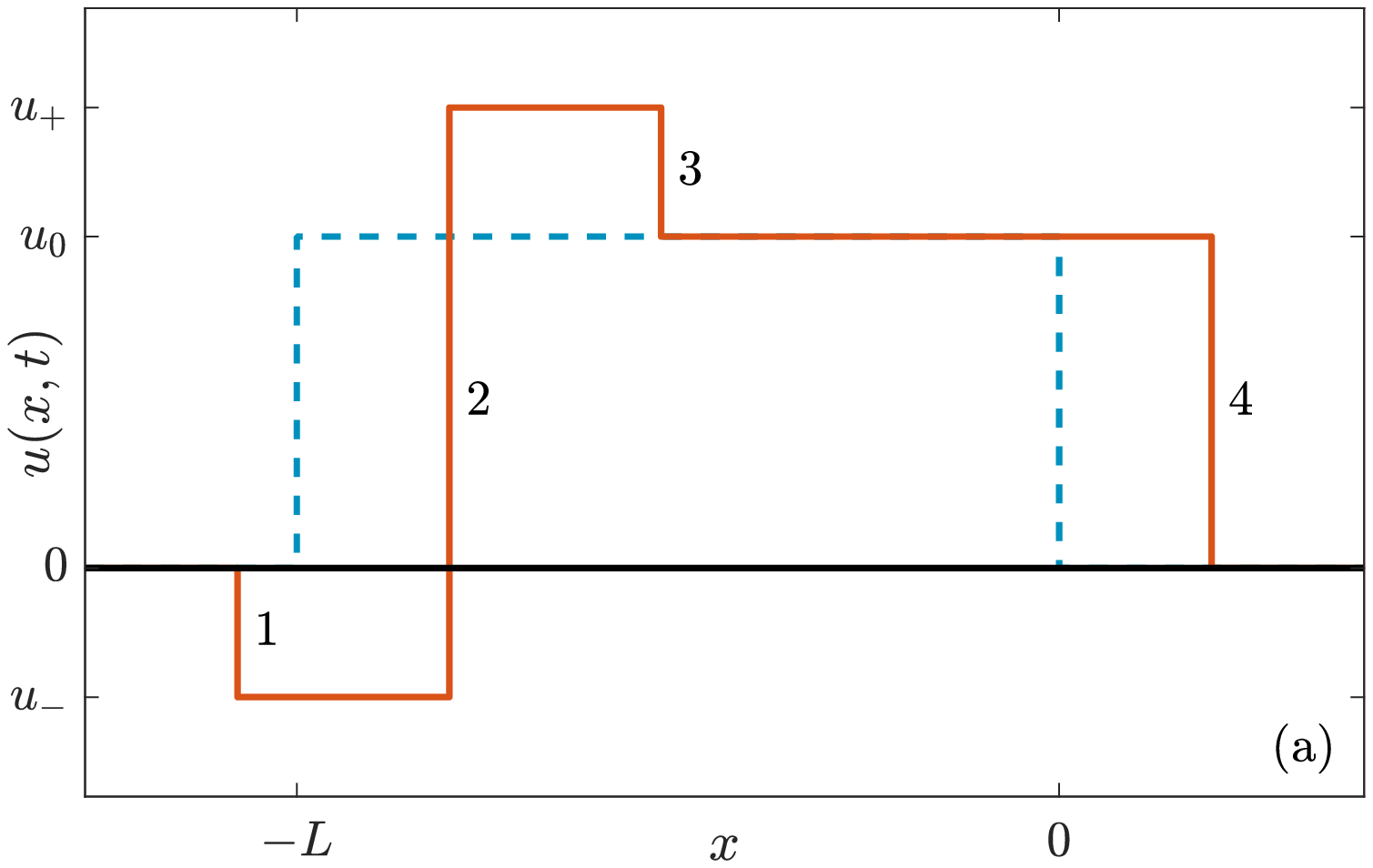}
\includegraphics[width=0.45\textwidth,clip=]{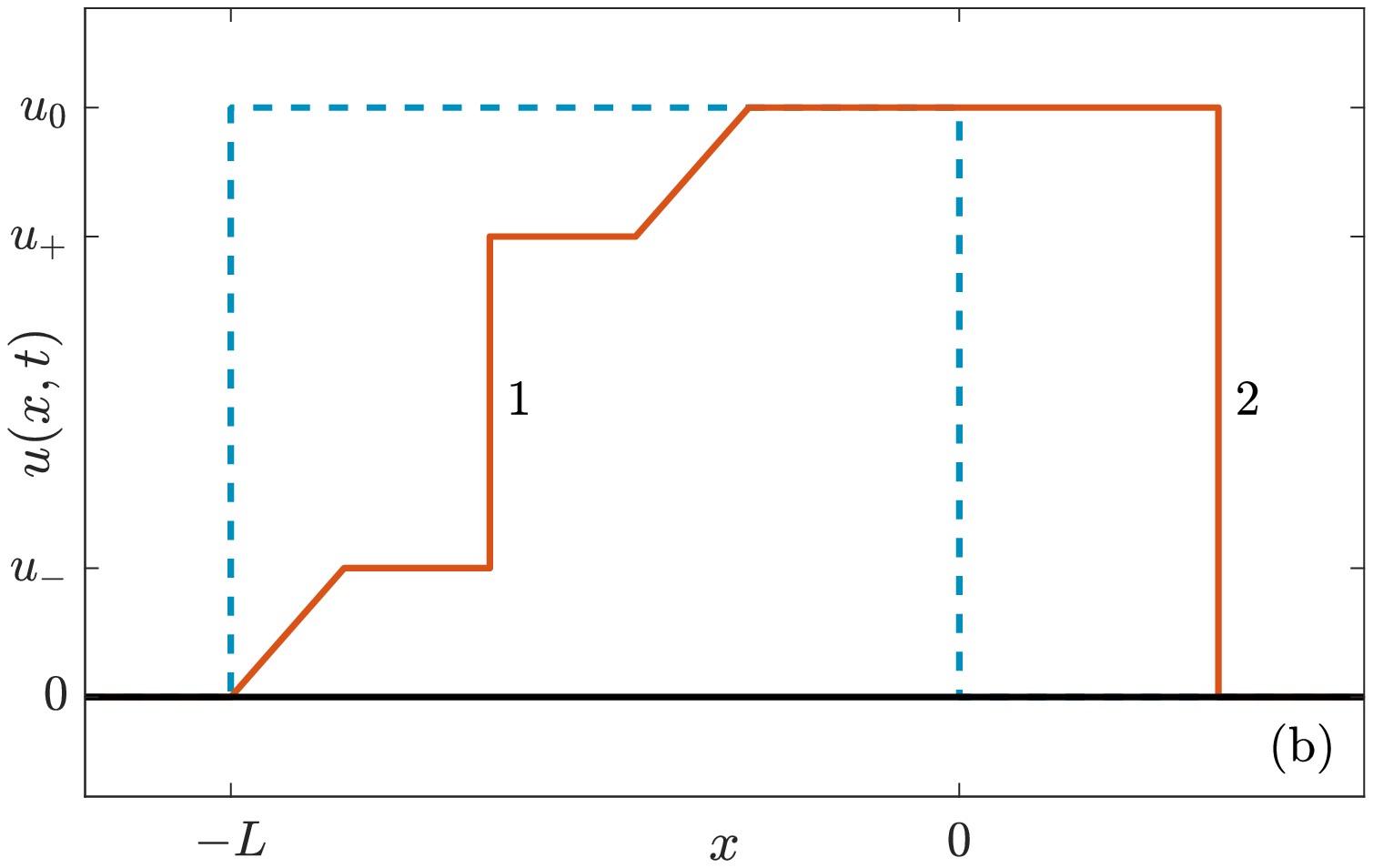}
\caption{The optimal path of the interface conditioned on reaching a very large negative height $h(x=0,t=1)=H$. The solid lines show the spatial profiles of the interface slope $u(x,t)=\partial_x h(x,t)$  at an intermediate time for $\sigma<1$ (a) and $\sigma>1$ (b). The initial profiles are shown by the dashed lines. The $u$-shocks 1, 3 and 4 (a) and 2 (b) are the ordinary shocks of the Hopf equation (\ref{eq:hopf}). The $u$-shocks 2 (a) and 1 (b) are fluctuation-driven shocks, or anti-shocks: they are described by the traveling soliton-shock solution (\ref{eq:rhosol})-(\ref{eq:Vsol}). At the hydrodynamic length scale the $p$-soliton is a point-like object which location coincides with that of the $u$-shock 2 (a) or 1 (b). The quantities $u_0$, $u_-$, $u_+$ and $L$ are given in terms of $\Lambda$ and $\sigma$ by Eqs.~(\ref{eq:v0}) and~(\ref{eq:v1v2L}) The shocks 3 and 4 (a) merge at time $t_m=2\sigma^2/(1+\sigma^2)<1$, but the merger occurs at $x>0$ and does not affect $h(0,1)$.}
\label{fig:vxt}
\end{figure}

As in the case of $\sigma=1$ \cite{JKM}, at very large negative $H$, the \emph{optimal initial condition} involves a large-amplitude narrow negative pulse of $p$, located at $x= -L$, where a negative plateau of $h$ becomes a ``ramp",
see Fig. \ref{fig:initial}. Correspondingly, the initial interface slope $u\simeq u_0 =\text{const}$ for $-L<x<0$
and zero elsewhere,  as indicated by the dashed lines in Fig. \ref{fig:vxt} a and b.  The parameters $L$ and $u_0$, and the plateau value $H_0=-L u_0$ are to be determined, alongside with other attributes of the solution.
At large $|H|$ the characteristic widths of the  $p$-pulse and of the $u$-jumps, which all scale as $1/\Delta u$, are small, see Eqs.~(\ref{eq:rhosol} and~(\ref{eq:Vsol}). (As we will ultimately see, these widths scale with $H$ as $1/\sqrt{|H||}\ll 1$.) In contrast, the ``hydrodynamic" length scale of the problem, which scales as $L \sim \sqrt{|H|}\gg 1$, is large. Therefore, at the hydrodynamic scale, the $p$-pulse and  the $u$-jumps can be treated as point-like objects, in a close analogy to fluid mechanics of compressible flow where, for example, shocks can be treated as discontinuities \cite{LL,Whitham}. In particular,  $p(x,t=0)$ is negligible in the vicinity of $x=0$, and the boundary condition~(\ref{eq:BC0}) yields
\begin{equation}\label{eq:v0}
    u_0 = - \frac{\Lambda \sigma^2}{2}.
\end{equation}
%##################################################################################%
%------------------------------------- FIG. 1 -------------------------------------%
%##################################################################################%
%##################################################################################%

%##################################################################################%
%------------------------------------- FIG. 2 -------------------------------------%
%##################################################################################%
%##################################################################################%

Here is how the optimal fluctuation develops in time.  The $u$-jump, which is initially at $x=0$, travels to the right, with velocity $u_0/2$, as a shock of the Burgers equation
\begin{equation}\label{eq:burgers}
\partial_t u+ u \partial_x u=\partial_{x}^2 u.
\end{equation}
At the hydrodynamic length scale the shock can be described by the inviscid Burgers equation, also called the
Hopf equation \cite{Whitham}:
\begin{equation}\label{eq:hopf}
\partial_t u+ u \partial_x u=0 .
\end{equation}
The $u$-jump, which is initially at $x= -L$, in general splits  into several ``elementary excitations". As we will see shortly, for $\sigma<1$
these elementary excitations are three distinct $u$-shocks. They are indicated by the numbers 1, 2 and 3 in Fig.~\ref{fig:vxt}a, whereas the shock initially at $x=0$ is indicated by number 4. The fluctuation-driven $u$-shock (\ref{eq:Vsol}), accompanied by the $p$-soliton (\ref{eq:rhosol}), is the shock number 2.

The $u$-shock 1 has $u=0$ on its left and $u=u_{-}$ on its right. The fluctuation-driven $u$-shock 2 has $u=u_{-}$ on its left and $u=u_{+}$ on its right. The $u$-shock 3 has $u=u_{+}$ on its left and $u=u_{0}$ on its right. Like the $u$-shock 4, the $u$-shocks 1 and 3 are ordinary shocks of the Hopf equation~(\ref{eq:hopf}). The $p$-soliton (\ref{eq:rhosol}), which accompanies the fluctuation-driven $u$-shock 2, rapidly emerges from the initial $p$-pulse at $x=-L$.

For  $\sigma>1$ the $u$-jump, initially at $x=-L$,
splits into a single $u$-shock and two so-called rarefaction waves \cite{LL,Whitham}, each of them obeying the equation $u(x,t)=(x+L)/t$, see Fig.~\ref{fig:vxt}b. The $u$-shock is
fluctuation-driven; it is accompanied by the $p$-soliton as described by Eqs.~(\ref{eq:rhosol}) and ~(\ref{eq:Vsol}).

The reasons for this flow structure become clear once the constants $u_-$, $u_+$ and $L$ are expressed via $\Lambda$ and $\sigma$. For that we need three relations. One of them is the conservation law $\int_{-\infty}^{\infty} p(x,t) \,dx =\Lambda$. When applied
to Eq.~(\ref{eq:rhosol}), it yields
\begin{equation}\label{eq:1}
u_{-}-u_{+}=\frac{\Lambda}{2}.
\end{equation}
An additional conservation law  is
\begin{equation}\label{eq:cons2}
\int_{-\infty}^{\infty} p(x,t) u(x,t)\,dx =\text{const}.
\end{equation}
The constant on the right hand side can be evaluated at $t=0$ by using Eq.~(\ref{eq:BC0}) and taking into account the fact that $p(x,0)$ is strongly localized around $x=-L$. As a result, the constant is equal to
$-u_0^2/\sigma^2$. Now, evaluating the left hand side of Eq.~(\ref{eq:cons2}) on the soliton-shock solution  (\ref{eq:rhosol}) and (\ref{eq:Vsol}), we obtain
\begin{equation}\label{eq:2}
(u_{-}-u_{+})(u_{-}+u_{+})=-\frac{u_0^2}{\sigma^2}.
\end{equation}
Finally, at $t=1$ the  $p$-soliton, which travels with the velocity $c$ from Eq.~(\ref{eq:c}), must arrive at $x=0$ \cite{delta}. This yields a kinematic condition,
\begin{equation}\label{eq:3}
\frac{ u_{-}+u_{+}}{2}=L.
\end{equation}
Equations~(\ref{eq:1}), (\ref{eq:2}) and (\ref{eq:3}) yield
\begin{equation}\label{eq:v1v2L}
    u_{-}=-\frac{\Lambda}{4} \left(\sigma^2-1\right) , \quad u_{+}=-\frac{\Lambda}{4} \left(\sigma^2+1\right),
    \quad L=-\frac{\Lambda  \sigma ^2}{4}\,.
\end{equation}
Notice that $\Lambda<0$ in the negative tail of $H$ that we are studying. The signs of, and the relation between, the quantities $u_-$, $u_+$ and $u_+$ determine the flow structure presented in Fig. \ref{fig:vxt} a and b \cite{Whitham}.

Using Eqs.~(\ref{eq:v0}) and (\ref{eq:v1v2L}), we can evaluate the action $s$ in terms of $\Lambda$ and $\sigma$. Indeed,
Eqs.~(\ref{eq:s0}) and~(\ref{eq:action1}) yield
\begin{equation*}
s_{\text{dyn}} \simeq \frac{1}{2}\times 1 \times \int_{-\infty}^{\infty} d\xi\,p^2(\xi)^2= \frac{1}{6} (u_{+}-u_{-})^3 = -\frac{\Lambda^3}{48}
\end{equation*}
and
\begin{equation*}
 s_{\text{in}} = \frac{u_0^2 L }{\sigma^2} =
-\frac{\Lambda ^3 \sigma^4}{16} ,
\end{equation*}
respectively. Here $s_{dyn}$ comes from the $p$-soliton (\ref{eq:rhosol}). As a result,
\begin{equation}\label{eq:svsLambda}
s=s_{\text{dyn}}+ s_{\text{in}} = -\frac{\Lambda ^3}{48}  \left(3 \sigma ^4+1\right).
\end{equation}
To express $\Lambda$ in terms of $H$ and $\sigma$ we use the condition~(\ref{eq:conditionH}) together with Eq.~(\ref{eq:h}) at $x=0$, where the terms $\partial_x^2 h$ and $p$ can be neglected. This yields
\begin{equation}\label{eq:h0}
h(x=0,t=1) \simeq -\frac{1}{2}\int_0^1  u^2(x=0,t)\,dt , %, = H.
\end{equation}
which should be equal to $H$. The calculations of $u(x=0,t)$ are different for $\sigma<1$ and $\sigma>1$. For $\sigma<1$ we have $u_{-}<0$, therefore the $u$-shock 1 (see Fig.~\ref{fig:vxt}) travels to the left and is inconsequential for our calculation. As  $u_{+}>u_0$, the excess $u_{+}-u_0$ travels to the right as the shock 3.
As a result,
\begin{numcases}
{u(x=0,t) =} u_0, & $0<t<\tau$,\nonumber \\
u_{+}, &$\tau<t<1$, \label{eq:v(0)}
\end{numcases}
where $\tau = 2 \sigma^2 (3\sigma^2+1)^{-1}<1$ is the time it takes  the shock 3 and the $p$-soliton to travel from $x=-L$ to the origin. Then Eqs.~(\ref{eq:h0}) and~(\ref{eq:v(0)}) yield
\begin{equation}\label{eq:height1}
 h(0,1) \simeq  -\frac{1}{2} u_0^2 \tau - \frac{1}{2} u_{+}^2 (1-\tau)
= -\frac{\Lambda ^2 }{32}\left(3 \sigma ^4+1\right).
\end{equation}
Equating this expression to $H$, we obtain
\begin{equation}\label{eq:Lambdaless}
\Lambda=- \left(\frac{32\left|H\right|}{3 \sigma ^4+1}\right)^{1/2} ,
\end{equation}
and Eq.~(\ref{eq:svsLambda}) becomes
\begin{equation}\label{eq:sless}
s=\frac{8\sqrt{2} \,|H|^{3/2}}{3\sqrt{3 \sigma^4+1}} .
\end{equation}

\begin{figure}
\includegraphics[width=0.5\textwidth,clip=]{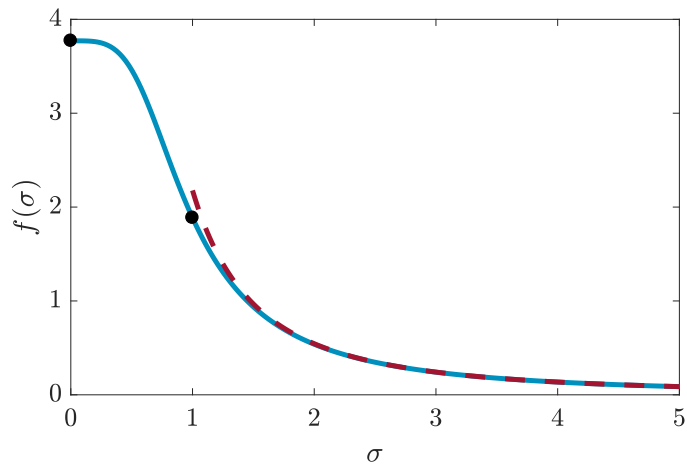}
\caption{The function $f(\sigma)$ which describes the $\lambda H>0$ tail, see Eqs.~(\ref{eq:negativetailsigma}) and (\ref{eq:fsigma}). The filled circles, at the points $(0,\frac{8\sqrt{2}}{3})$ and $(1,\frac{4\sqrt{2}}{3})$, correspond to the flat and stationary initial data, respectively. The dashed line shows the large-$\sigma$ asymptote (\ref{eq:largesigma}).}
\label{fig:f}
\end{figure}

For $\sigma>1$
we have $u_{+}<u_0$. As a result, two rarefaction waves [obeying the same relation $u(x,t)=(x+L)/t$] develop on the left and on the right of the fluctuation-driven shock. There are also two weak discontinuities,
see Fig.~\ref{fig:vxt}b.
The spatial arrangement of the fluctuation-driven shock and the rarefaction waves do not change until $t=1$. Therefore, the left rarefaction wave is inconsequential for the purposes of evaluating $h(0,1)$. In the relevant regions of space we have
\begin{numcases}
{u(x,t) =}u_{+}, & $c t <x+L<u_{+}t$,\label{eq:left} \\
\frac{x+L}{t}, &$u_{+}t<x+L<u_{0} t$, \label{eq:rarefaction}\\
u_0, &$u_{0} t<x+L<u_0 t/2+L$. \label{eq:right}
\end{numcases}
As a result,
\begin{numcases}
{u(0,t) =}u_{0}, & $0<t<\tau_1$,\label{eq:stage1} \\
L/t, &$\tau_1<t<\tau_2$, \label{eq:stage2}\\
u_{+}, &$\tau_2<t<1$, \label{eq:stage3}
\end{numcases}
where $\tau_1=L/u_0=1/2$ and $\tau_2=L/u_{+}=\sigma^2/(1+\sigma^2)$.
Using Eq.~(\ref{eq:h0}) and Eqs.~(\ref{eq:stage1})-(\ref{eq:stage3}), we obtain
\begin{equation}\label{eq:h01more}
h(0,1)=-\frac{\Lambda^2}{32}  \left(3 \sigma ^4+1\right).
\end{equation}
Somewhat surprisingly, this formula coincides with Eq.~(\ref{eq:height1}), obtained for $\sigma<1$. As a result, $\Lambda$ versus $H$ is still described by Eq.~(\ref{eq:Lambdaless}), and the action~(\ref{eq:svsLambda}) is described by Eq.~(\ref{eq:sless}) for \emph{all} $\sigma$. Equation~(\ref{eq:sless}) brings us to the main analytic result of this work, announced in Eq.~(\ref{eq:negativetailsigma}), where
\begin{equation}
f(\sigma) = \frac{8\sqrt{2}}{3\sqrt{3 \sigma^4+1}}, \label{eq:fsigma}
\end{equation}
A graph of the function $f(\sigma)$ is shown in Fig. \ref{fig:f}. For $\sigma=1$ we obtain $f=(4/3)\sqrt{2}$, reproducing the BR distribution tail which is observed, for the stationary initial condition, at all times \cite{IS,Borodinetal,JKM,KD}.
%##################################################################################%
%------------------------------------- FIG. 3 -------------------------------------%
%##################################################################################%
%##################################################################################%

For $\sigma=0$ we obtain $f= (8/3)\sqrt{2}$. This describes the $\lambda H\gg 1$ tail of
the GOE Tracy-Widom distribution which is observed, at all times,  for flat initial condition \cite{KK2007,CLD,MKV}.
Indeed, at $\sigma \to 0$, the fluctuational cost~(\ref{eq:s0}) of a non-flat interface at $t=0$ becomes prohibitively high.
As a result, the optimal initial interface is almost flat,
and the leading-order contribution to the action mostly comes from the dynamics.

\begin{figure}
\includegraphics[height=0.3\textwidth,clip=]{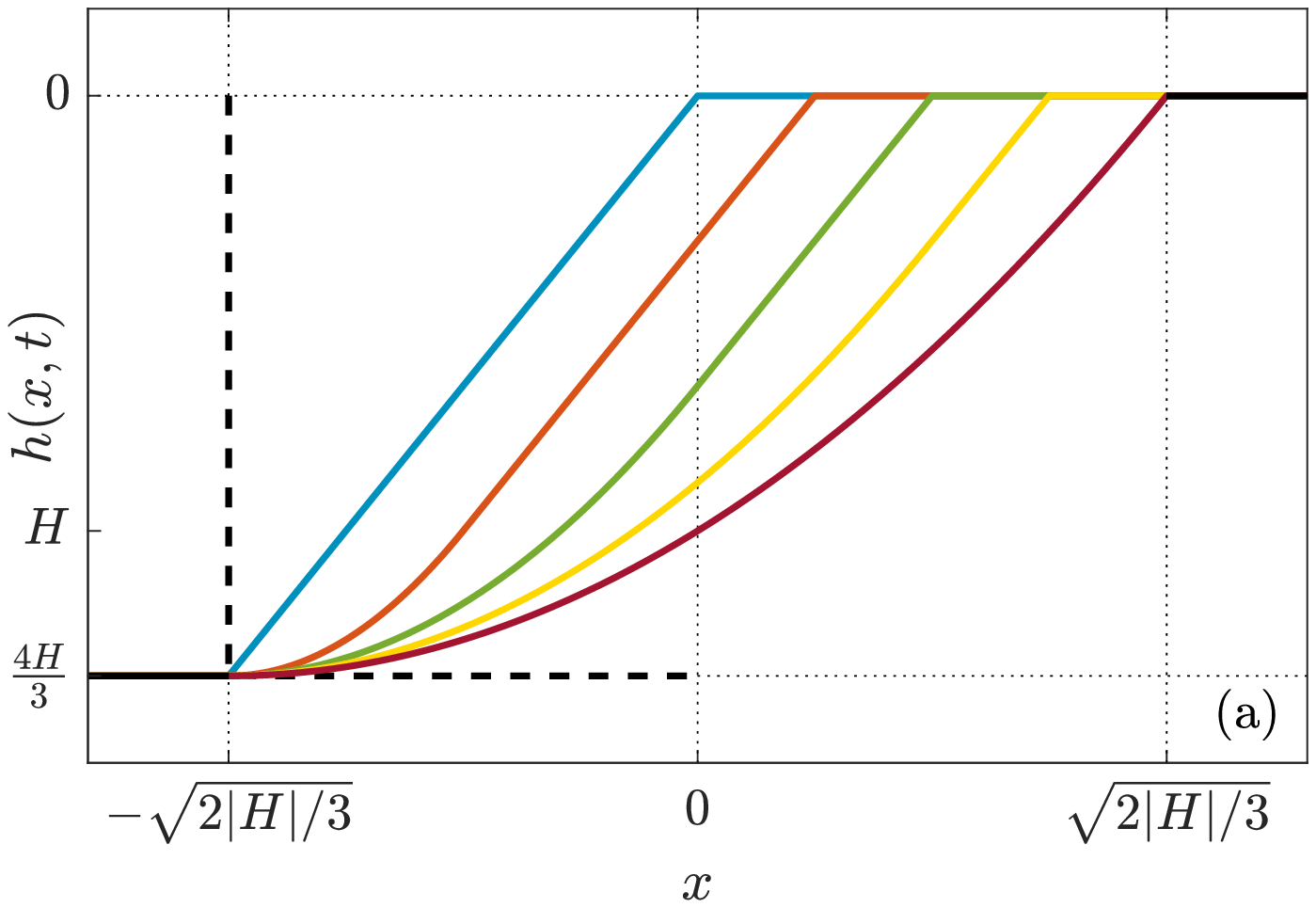}
\includegraphics[height=0.3\textwidth,clip=]{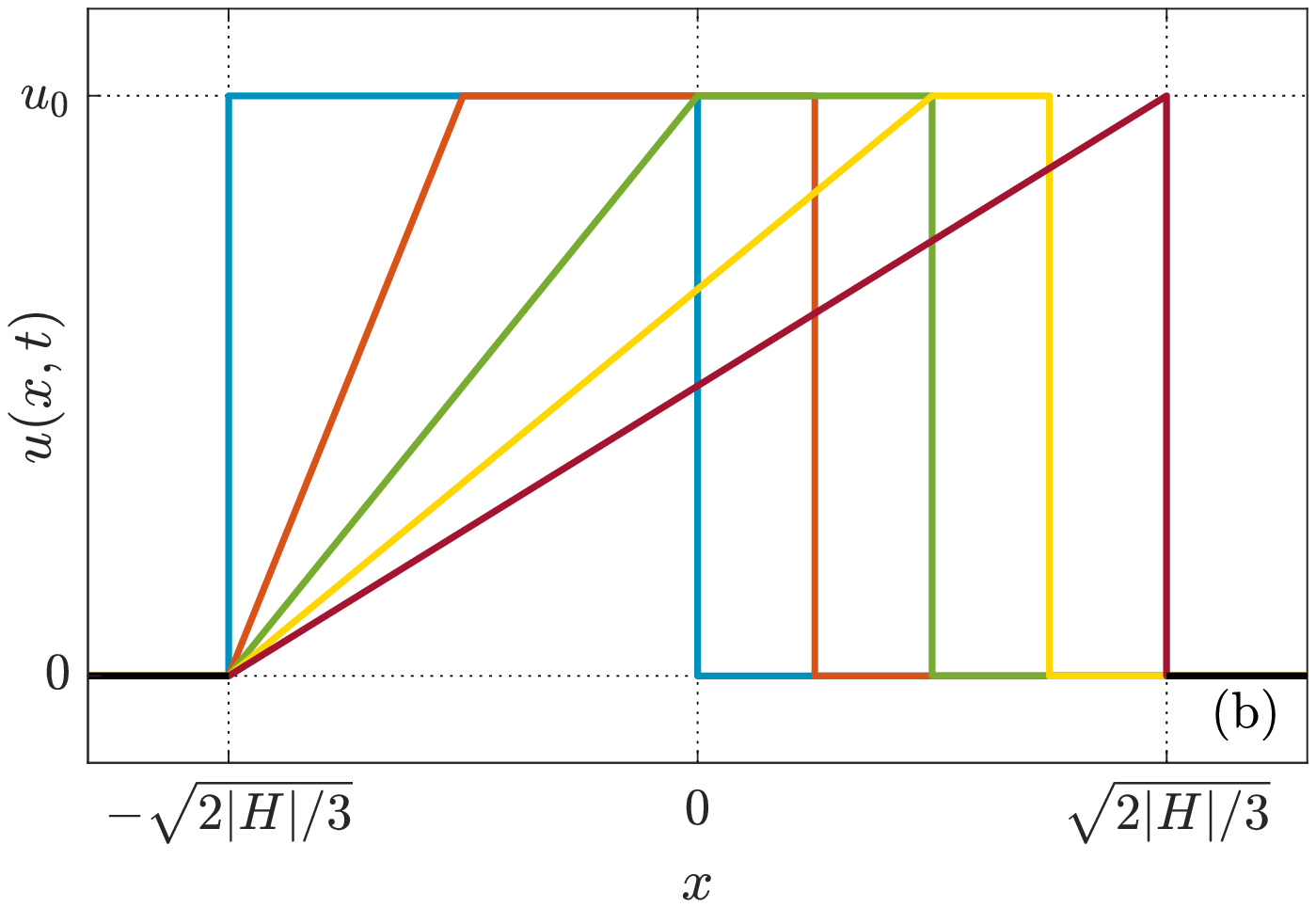}
\caption{The optimal interface profile $h(x,t)$  and slope $u(x,t)$ in the limit of $-H\gg \sigma^2 \gg 1$. Left to right: the optimal initial $h$-profile ($u$-slope) and the $h$-profiles ($u$-slopes) at times $ 1/4, 1/2, 3/4$ and $1$ governed by the Hopf equation (\ref{eq:hopf}).}
\label{fig:relax}
\end{figure}

The limit of $\sigma \gg 1$ is also interesting. Here the leading-order asymptote,
\begin{equation}\label{eq:largesigma}
f(\sigma) \simeq \frac{8\sqrt{2}}{3\sqrt{3} \,\sigma^2} ,
\end{equation}
comes from the (very low) cost $s_{\text{in}}$
of a nontrivial optimal initial condition, with $u_0 = 2 \sqrt{2/3}\,|H|^{1/2}\gg 1$ and $L=\sqrt{2/3}\,|H|^{1/2} \gg 1$.  Figure \ref{fig:relax} shows the corresponding optimal optimal path of $h(x,t)$. To leading order, this optimal path is fluctuationless, $p(x,t) = 0$; it describes \emph{relaxation} of the optimal initial interface. The relaxation is  governed by the Hopf equation (\ref{eq:hopf}). Notice that the plateau height $|H_0|=(4/3)|H|$ is larger than $|H|$. Actually, the inequality $|H_0|>|H|$ holds for all $\sigma>1$. For $\sigma<1$ one has $|H_0|<|H|$. The equality $|H_0|=|H|$ holds only for $\sigma=1$. These properties follow from the expression
\begin{equation}\label{H0general}
H_0= -\frac{4\sigma^4 \,|H| }{3 \sigma^4+1},
\end{equation}
valid for any $\sigma$.

The results of this subsection hold if the characteristic soliton/shock width is much smaller that
the hydrodynamic length scale $L$, as we assumed. For $\sigma<1$ this condition demands $|H|\gg 1$, whereas for $\sigma\gg 1$ it is more stringent: $|H|\gg \sigma^2$.

%##################################################################################%
%------------------------------------- FIG. 4 -------------------------------------%
%##################################################################################%
%##################################################################################%

Although the function $f(\sigma)$ does not have any singularity at $\sigma=1$, the stationary case $\sigma=1$ is special
in terms of the optimal path. Here the optimal initial $u$-jump
does \emph{not} split into multiple shocks, neither does it produce rarefaction waves. Instead, the $p$-pulse and the $u$-jump directly, and rapidly evolve into a traveling soliton-shock pair (\ref{eq:rhosol}) and (\ref{eq:Vsol}) with $u_{+}=u_0$ and $u_{-}=0$, as was observed in Ref. \cite{JKM}. This simple behavior appears because, at $\sigma=1$, the initial condition (\ref{eq:BC0}) obeys Eq.~(\ref{eq:invariant}) everywhere except at the origin, where the delta-function appears. The ordinary $u$-shock at $x=0$ ``takes care" of the delta, and the
ordinary shock and the fluctuation-driven shock travel with the same velocity, creating the ``shock-antishock" pair
\cite{Fogedby1998,Fogedby1999}, see also Refs. \cite{BD2005,BD2006}.

When $\sigma$ is different from $1$, the initial condition (\ref{eq:BC0}) violates Eq.~(\ref{eq:invariant}) everywhere. As a result, additional $u$-shocks with different velocities (for $\sigma<1$), or rarefaction waves (for $\sigma>1$), emerge to cure this violation. It is surprising that the different structure of solutions for $\sigma<1$ and $\sigma>1$ does not cause different functional dependences  $\Lambda=\Lambda(H)$ for $\sigma<1$ and $\sigma>1$, and a non-smoothness of $f(\sigma)$ at $\sigma=1$.

In the dimensional units of Eq.~(\ref{eq:KPZoriginal}) the $\lambda H>0$ tail is as announced
in Eq.~(\ref{eq:negativetailsigma}).

%&&&&&&&&&&&&&&&&&&&&&&&&&&&&&&&&&&&&&&&&&&&&&&&&&&&&&&&&&&&&&&&&&&&&&&&&&&&&&&&&&&&&&&&&&&&&&&&&&&&&&&&&&&&&&&&&&&&&&&&&&&&&&&&&&&&&&&
%&&&&&&&&&&&&&&&&&&&&&&&&&&&&&&&&&&&&&&&&&&&&&&&&&&&&&&&&&&&&&&&&&&&&&&&&&&&&&&&&&&&&&&&&&&&&&&&&&&&&&&&&&&&&&&&&&&&&&&&&&&&&&&&&&&&&&&
\subsection{$\lambda H<0$ large-deviation tail}
\label{sec:positive H}

The $\lambda H<0$ large-deviation tail has quite a different nature. Here the optimal path respects the reflection symmetry $x\leftrightarrow -x$ at any $H$. As in Refs. \cite{KK2009,MKV,KMS,JKM}, the leading-order approximation to the OFM solution can be obtained in the framework of ``inviscid hydrodynamics", that is by neglecting the diffusion terms in Eqs.~(\ref{eq:h}), (\ref{eq:Leqrho}) and (\ref{eq:BC0}). In this approximation the
``fluctuation cost" of the initial profile, $s_{\text{in}}$ from Eq.~(\ref{eq:s0}), is negligible. As a result,
the initial condition (\ref{eq:BC0}) becomes simply $p(x,t=0)=\Lambda \delta(x)$, and
the parameter $\sigma$ drops out of the problem.  This case was solved in Refs. \cite{KMS,JKM}.
The resulting  $\lambda H<0$ large deviation tail coincides, in the leading order, with the corresponding tail
for the (deterministic) droplet initial condition: $s\simeq s_{\text{dyn}} = 4\sqrt{2} H^{5/2}/(15 \pi)$
\cite{KMS,DMRS,SMP,KD}. In dimensional units
\begin{equation}\label{eq:linearcond}
-\ln \mathcal{P}_t(H,\sigma)\simeq \frac{4\sqrt{2 |\lambda|}}{15 \pi D}\, \frac{H^{5/2}}{t^{1/2}}.
\end{equation}
Therefore, at fixed $\sigma>0$, and for sufficiently large $H$,  the $\lambda H<0$ large-deviation tail
is, in the leading order, independent of $\nu$ and $\sigma$. The optimal path of the
interface is different from that for the droplet: it includes a fluctuationless $u$-shock with a non-trivial dynamics
\cite{JKM}.

%&&&&&&&&&&&&&&&&&&&&&&&&&&&&&&&&&&&&&&&&&&&&&&&&&&&&&&&&&&&&&&&&&&&&&&&&&&&&&&&&&&&&&&&&&&&&&&&&&&&&&&&&&&&&&&&&&&&&&&&&&&&&&&&&&&&&&&
%&&&&&&&&&&&&&&&&&&&&&&&&&&&&&&&&&&&&&&&&&&&&&&&&&&&&&&&&&&&&&&&&&&&&&&&&&&&&&&&&&&&&&&&&&&&&&&&&&&&&&&&&&&&&&&&&&&&&&&&&&&&&&&&&&&&&&&
\subsection{$\mathcal{P}_t(H,\sigma)$ at short and long times}
\label{sec:longtime}

As we have seen, at short times the (Gaussian) body of the distributions $\mathcal{P}_t(H,\sigma)$ is described by Eq.~(\ref{eq:variance}),
while the $\lambda H>0$ and $\lambda H<0$ tails are described by Eqs.~(\ref{eq:negativetailsigma}) and (\ref{eq:linearcond}), respectively.

At long times the Gaussian distributions give way to the non-Gaussian distributions observed in Ref. \cite{CFS}. As the parameter $\epsilon \sim \sqrt{t}$ is now large, the OFM does not apply to typical fluctuations.  However, it continues to apply sufficiently far in the tails. It has been shown that the $H^{3/2}/t^{1/2}$ tails of the GUE and GOE TW distribution (for
the droplet and flat interface), and the similar $H^{3/2}/t^{1/2}$ tail of the BR distribution,
persist at \emph{all} times \cite{KMS,KK2007,DMRS,DMS,MKV,JKM,KD} and are correctly described by the OFM. Therefore, it is natural to assume that the same is true for
the more general random initial conditions considered in Ref. \cite{CFS} and in the present work. That is, one should expect that the leading-order asymptotic~(\ref{eq:negativetailsigma}) and (\ref{eq:fsigma}) provides a correct long-time description of the  $\lambda H>0$ tail of $\mathcal{P}_t(H,\sigma)$. In the next section we put this assumption into test by performing large-scale MC simulations of the TASEP.

The other distribution tail, $\lambda H<0$, has a more complicated structure at long times. Here a different tail appears at moderately large heights and ``pushes" the large-deviation tail (\ref{eq:linearcond}) toward progressively larger heights as time increases.  For the three basic initial conditions this emerging non-OFM tail behaves as $H^3/t$: it is the tail of the corresponding TW/BR distribution.  The ``inviscid" large-deviation tail (\ref{eq:linearcond}), however, is still present at $|H|\gg t\gg 1$. The ensuing dynamical two-tail structure was conjectured for a whole class of initial conditions in Refs. \cite{MKV,KMS,JKM}. Recently the two-tail structure has been obtained analytically, and explicitly, for the droplet case \cite{SMP} by using an exact representation for $\mathcal{P}_t(H)$ derived in Ref. \cite{ACQ}. The crossover between the inner $H^3/T$ tail and outer $H^{5/2}/T^{1/2}$ tail occurs at $H\sim t$. It is natural to assume that, at long times, the $\lambda H<0$ tail of $\mathcal{P}_t(H,\sigma)$  has a similar dynamical two-tail structure for the general Brownian initial conditions. We also note that the large-deviation tail (\ref{eq:linearcond}) may be specific for the KPZ equation.

%#$#$#$#$#$#$#$#$#$#$#$#$#$#$#$#$#$#$#$#$#$#$#$#$#$#$#$#$#$#$#$#$#$#$#$#$#$#$#$#$#$#$#$#$#$#$#$#$#$#$#$#$#$#$#$#$#$#$#$#$#$#$#$#$#$#$#$
%#$#$#$#$#$#$#$#$#$#$#$#$#$#$#$#$#$#$#$#$#$#$#$#$#$#$#$#$#$#$#$#$#$#$#$#$#$#$#$#$#$#$#$#$#$#$#$#$#$#$#$#$#$#$#$#$#$#$#$#$#$#$#$#$#$#$#$
%#$#$#$#$#$#$#$#$#$#$#$#$#$#$#$#$#$#$#$#$#$#$#$#$#$#$#$#$#$#$#$#$#$#$#$#$#$#$#$#$#$#$#$#$#$#$#$#$#$#$#$#$#$#$#$#$#$#$#$#$#$#$#$#$#$#$#$
\section{Monte-Carlo simulations of the TASEP with Brownian initial conditions}
\label{sec:TASEP}
We tested our analytical predictions for the $\lambda H>0$ tail in MC simulations. In Appendix A we employ Nonlinear Fluctuating Hydrodynamics (NLFH) to establish a duality between the TASEP and the KPZ equation. In Appendix B we present analytical expressions, for the parallel-update TASEP, for the quantities that appear in the NLFH description. Using the TASEP-KPZ duality and the analytical expressions, we accessed the height distributions $\mathcal{P}_t(H,\sigma)$ of the KPZ interface by simulating a TASEP with particle density $\rho=1/2$ and parallel update rule \cite{JS:Schadneider_STCS_book_2011}. In a parallel-update TASEP all particles attempt to hop to the left at the same time with probability $q\in(0,1)$, see Fig.~\ref{fig:JS:TASEP_growth_map} in Appendix A. The move is disallowed if the target site is occupied.

Evolving the TASEP configuration in time is a major bottleneck in terms of the computer memory and speed.
Pre-caching techniques allow to overcome this bottleneck for a sequential memory readout. Therefore, using a parallel (sequential read) instead of a random sequential (random read) update is crucial for reducing the computation time. Another problem is caused by the fact that
the initial configurations of the TASEP become more and more clustered with an increase of $\sigma$. For large $\sigma$ the clusters become a major obstacle to reaching the long-time asymptotic statistics, so they need to be dissolved.
For a random sequential update the cluster dissolution rate is constant,
while for a parallel update it can be increased by increasing the hopping probability $q$. Note that the random sequential update is recovered in the limit of $q\rightarrow 0$ (when time is properly rescaled), while in the limit of $q\rightarrow 1$ the dynamics become deterministic. In our simulations we used the hopping probability $q=1/2$ for $0\leq\sigma\leq1$, and $q=15/16$ for $\sigma\geq1$.

Our first objective was to access the late-time height (correspondingly, the time-integrated current) distribution which is expected to have the following scaling behavior:
\begin{equation}
\mathcal{P}_{t}\left(H ,\sigma\right)  \simeq
\left(\frac{\nu^{2}}{ D^{2}\left|\lambda\right| t}\right)^{1/3}
F^{\left(\sigma\right)}\left[-\left(\frac{\nu^{2}}{D^{2}\left|\lambda\right|t}\right)^{1/3} H\right] .
\label{eq:JS:Scaling_relation_Height_Fluc}
\end{equation}

\begin{figure}
\includegraphics[width=0.45\textwidth]{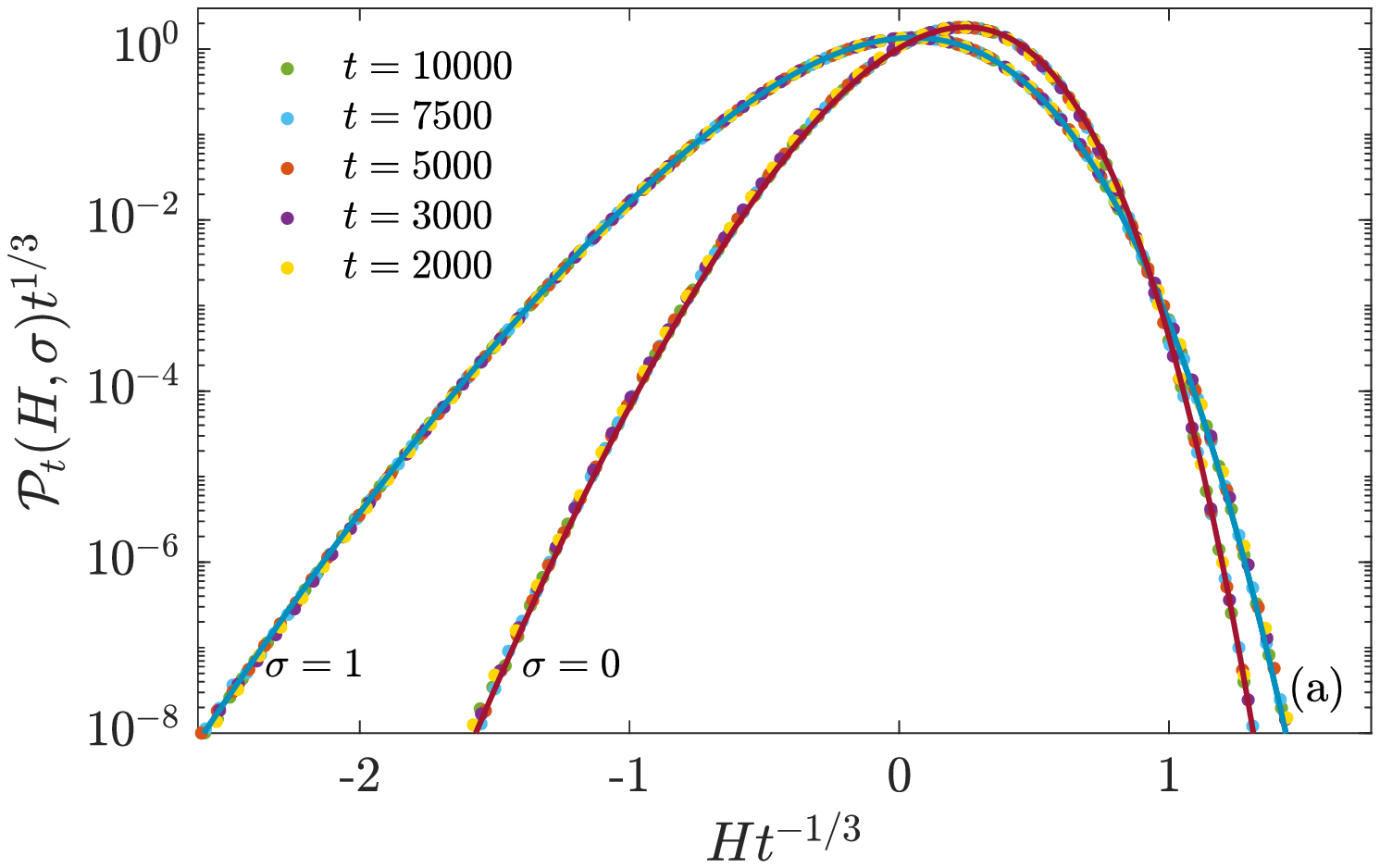}
\includegraphics[width=0.45\textwidth]{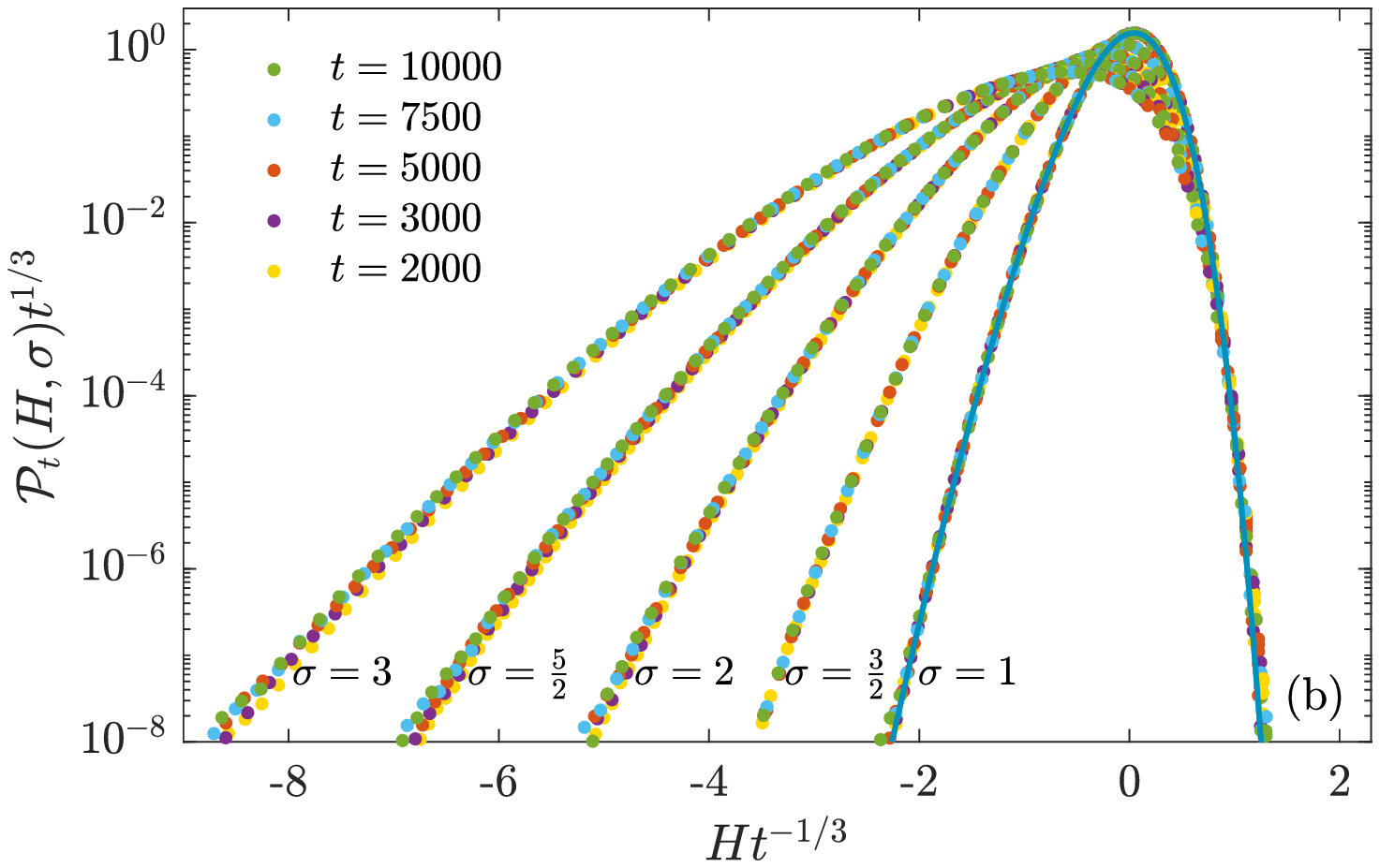}
\caption{Time collapse of height distributions for initial states with different $\sigma$. The height distributions are recorded from current fluctuations of a parallel-update TASEP with hopping probability $q=1/2$ (a) and $q=15/16$ (b). The TASEP current statistics are translated into height statistics using Eqs.~(\ref{eq:JS:FLucDissTheo}), ~(\ref{eq:JS:KPZ_NLFH_identities})~and~(\ref{eq:JS:Height_Current_Identity}). The solid lines are the asymptotic Baik-Rains ($\sigma=1$)  and GOE Tracy-Widom ($\sigma=0$) distributions for the stationary and flat initial conditions, respectively. The data match these distributions within statistical errors (of the order of symbol size) without adjustable parameters.}
\label{fig:JS:Time_Collapse_BR_GOE}
\end{figure}

To suppress finite-size corrections  of order $L^{-1}$ to the surface growth speed (correspondingly, to the time-integrated current), the  height/current statistics were recorded in periodic systems of size $L=10^9$. The statistics were extracted at every lattice site using the  translational invariance of the system. For each $\sigma$, the distributions were obtained from at least $500$ independent MC samples when starting from an initial state drawn from the $\sigma$-distribution described by Eqs.~(\ref{eq:JS:initial_configuration_distribution})~and~(\ref{JS:eq:sigma_2_cluster_prob}).
To verify that the asymptotic regime was reached, we tested the expected scaling behavior~(\ref{eq:JS:Scaling_relation_Height_Fluc}) at different times. As shown
in Fig.~\ref{fig:JS:Time_Collapse_BR_GOE} a very good time collapse is observed within statistical errors. The rescaled curves match, without any adjustable parameters, the BR and the GOE TW distributions for the stationary and flat initial conditions,
respectively.

\begin{figure}
\includegraphics[width=0.45\textwidth]{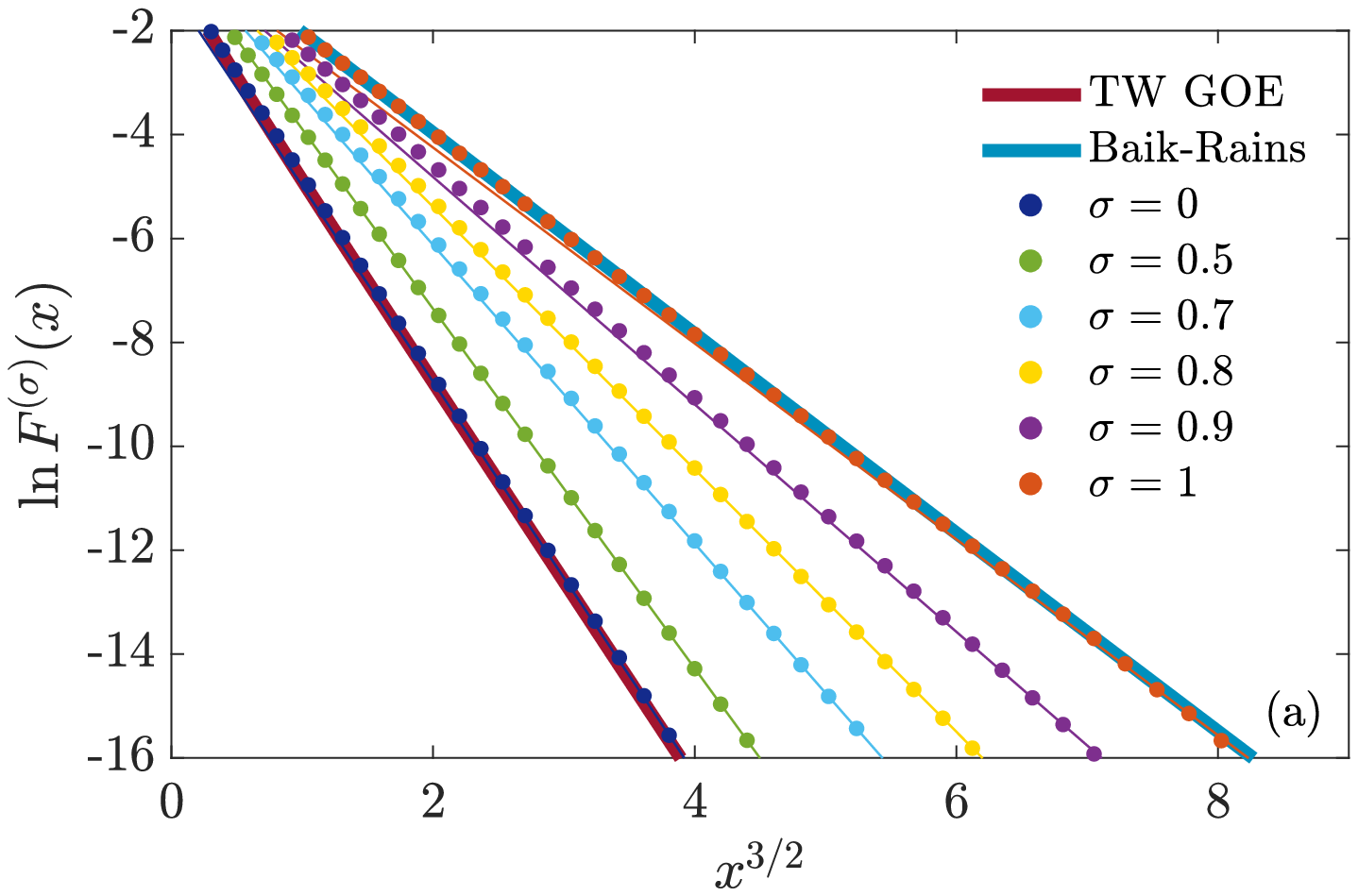}
\includegraphics[width=0.45\textwidth]{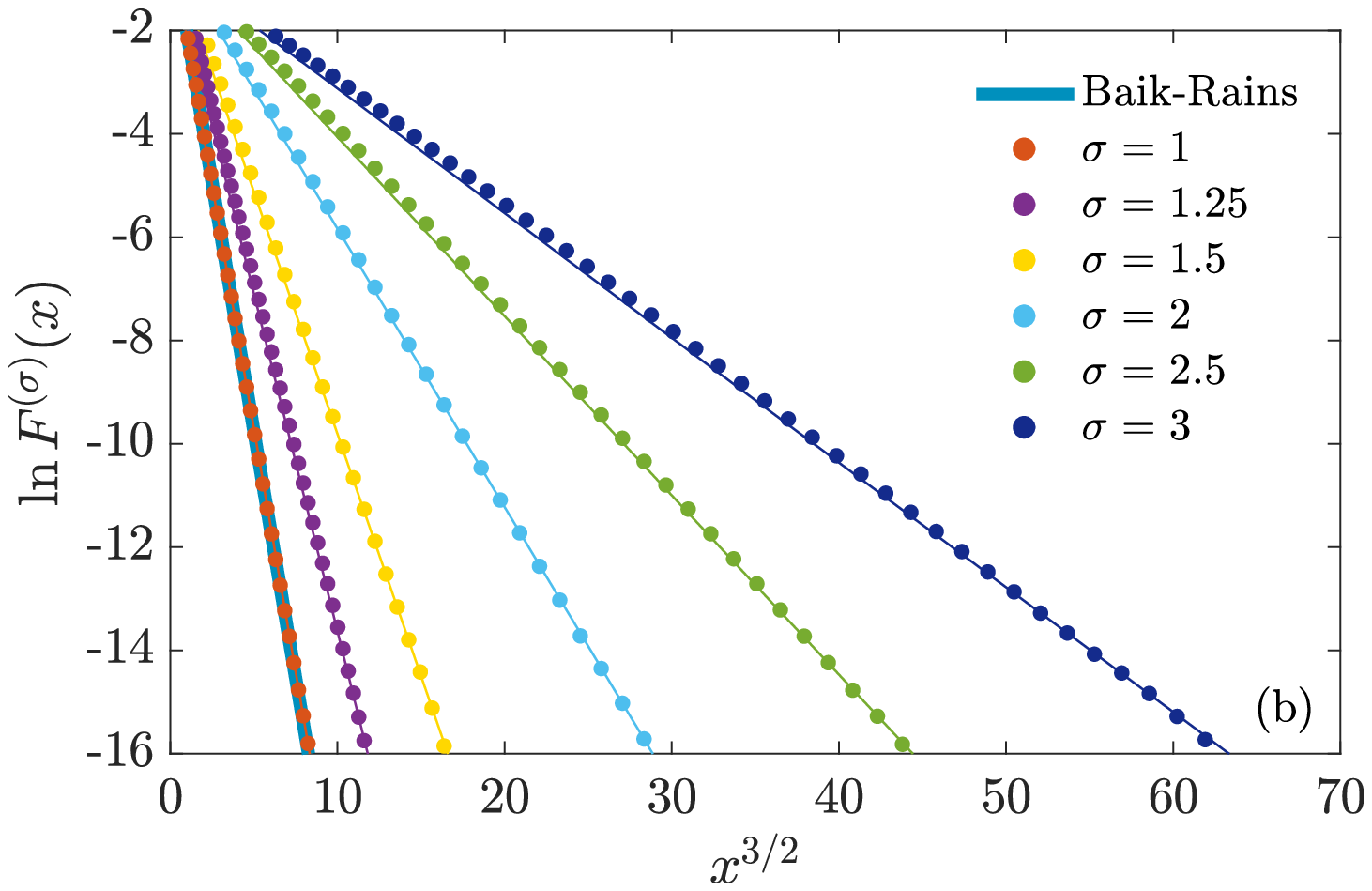}
\caption{$F^{(\sigma)}(x)$ calculated from Eq.~(\ref{eq:JS:Scaling_relation_Current_Fluc}) for different $0\leq\sigma\leq1$ (a) and $\sigma\geq1$ (b) from the TASEP current distribution. $\sigma$ increases from left to right. The current distribution was recorded at time $t=10,000$ using the hopping probabilities $q=1/2$ (a) and $q=15/16$ (b). The thick blue and red lines are the Baik-Rains and Tracy-Widom GOE distributions, respectively,  for $\sigma=1$ and $\sigma=0$.  The thin colored lines are a guide for the eye given by $y=-f(\sigma)  x^{3/2}+c_\sigma$. Statistical errors are of order of symbol size.}
\label{fig:JS:Slow_Tail_small_sigma}
\end{figure}

\begin{figure}
\includegraphics[width=0.45\textwidth]{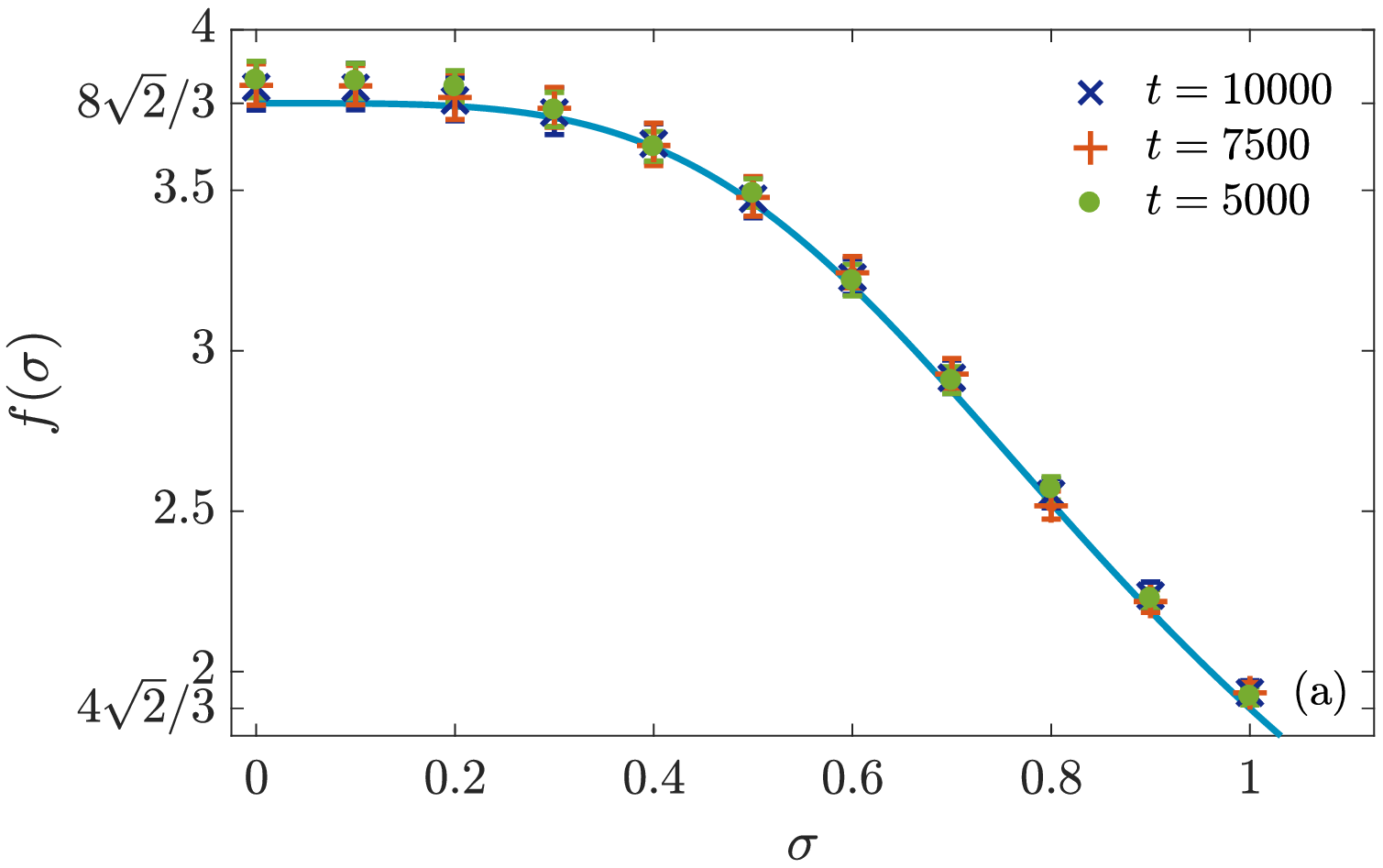}
\includegraphics[width=0.45\textwidth]{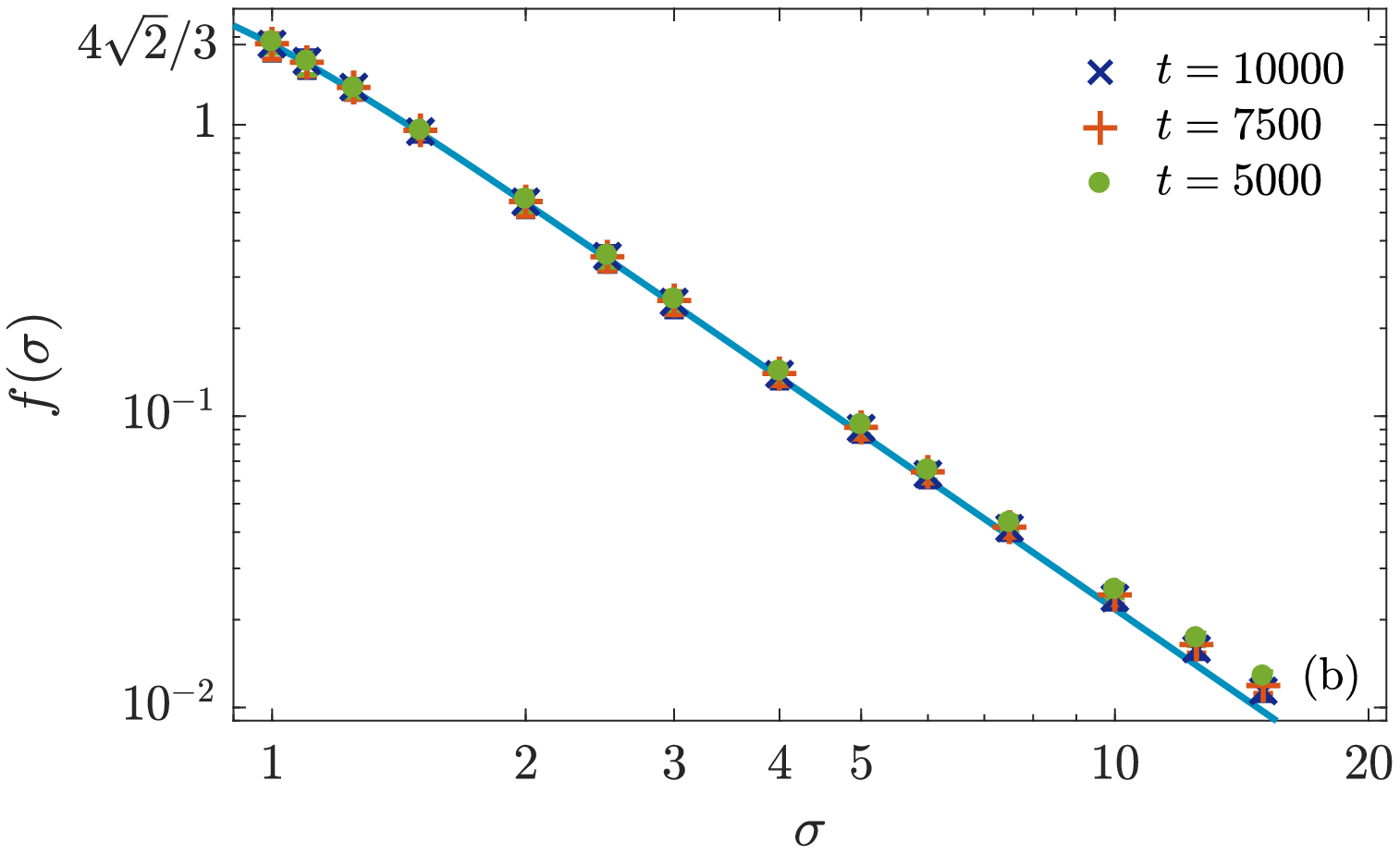}
\caption{The function $f(\sigma)$ from Eqs.~(\ref{eq:negativetailsigma}) and (\ref{eq:fsigma}) (the solid lines) is
compared to the measured slope of the slow tail of the TASEP current distribution, see Eq.~(\ref{eq:JS:Scaling_relation_Height_Fluc}) and Fig.~\ref{fig:JS:Slow_Tail_small_sigma}.
The current distribution was recorded for different $\sigma$  and and times $t$ using the hopping probabilities $q=1/2$ (a) and $q=15/16$ (b). Error bars indicate the $99\%$ confidence level.}
\label{fig:JS:f_sigma_fit_small_sigma}
\end{figure}

Having established the validity of Eq.~(\ref{eq:JS:Scaling_relation_Height_Fluc}), we  calculated $F^{(\sigma)}$ for different $\sigma$ and observation times $t$.  As can be seen from Fig.~\ref{fig:JS:Slow_Tail_small_sigma}, the measured MC data suffices to resolve the asymptote $\ln F^{(\sigma)}(x)\simeq f(\sigma)x^{3/2}$.
Finally, we fitted the slopes $f(\sigma)$ in the regime $8\leq-\ln F^{(\sigma)}(x)\leq 15$ and compared them to the analytical prediction, Eq.~(\ref{eq:fsigma}). As demonstrated in Fig.~\ref{fig:JS:f_sigma_fit_small_sigma}, the simulations and theory are in a very good agreement.

%##################################################################################%
%------------------------------------- FIG. 5 -------------------------------------%
%##################################################################################%
%##################################################################################%

%##################################################################################%
%------------------------------------- FIG. 6 -------------------------------------%
%##################################################################################%
%##################################################################################%

%##################################################################################%
%------------------------------------- FIG. 7 -------------------------------------%
%##################################################################################%
%##################################################################################%

\vspace{1cm}

%#$#$#$#$#$#$#$#$#$#$#$#$#$#$#$#$#$#$#$#$#$#$#$#$#$#$#$#$#$#$#$#$#$#$#$#$#$#$#$#$#$#$#$#$#$#$#$#$#$#$#$#$#$#$#$#$#$#$#$#$#$#$#$#$#$#$#$
%#$#$#$#$#$#$#$#$#$#$#$#$#$#$#$#$#$#$#$#$#$#$#$#$#$#$#$#$#$#$#$#$#$#$#$#$#$#$#$#$#$#$#$#$#$#$#$#$#$#$#$#$#$#$#$#$#$#$#$#$#$#$#$#$#$#$#$
%#$#$#$#$#$#$#$#$#$#$#$#$#$#$#$#$#$#$#$#$#$#$#$#$#$#$#$#$#$#$#$#$#$#$#$#$#$#$#$#$#$#$#$#$#$#$#$#$#$#$#$#$#$#$#$#$#$#$#$#$#$#$#$#$#$#$#$
\section{Summary}
\label{sec:discussion}
For a whole class of initial conditions, the $\lambda H>0$ tail of the one-point height distribution of the KPZ equation in $1 + 1$ dimensions persists at all times $t>0$. Here we exploited this persistence by using the Optimal Fluctuation Method, which is asymptotically exact in the limit of $t\to 0$. This enabled us to determine the optimal path of the conditioned process and calculate the $\lambda H>0$ tail analytically for a family of Brownian initial interfaces parameterized by their diffusion constants.   The $\lambda H>0$ tail is given by Eqs.~(\ref{eq:negativetailsigma}) and ~(\ref{eq:fsigma}). Our extensive MC simulations
with a parallel-update TASEP strongly support these results.

We also determined the  $\lambda H<0$ tail of the height distribution, Eq.~(\ref{eq:linearcond}). This $H^{5/2}/t^{1/2}$  tail is independent of the diffusion constant of the initial interface. By analogy with other initial conditions, we expect that at long times this tail is ``pushed" towards progressively larger heights.

In conclusion, our results show yet again that the OFM can be remarkably successful in the description of large fluctuations in
macroscopic systems far from equilibrium.

%#$#$#$#$#$#$#$#$#$#$#$#$#$#$#$#$#$#$#$#$#$#$#$#$#$#$#$#$#$#$#$#$#$#$#$#$#$#$#$#$#$#$#$#$#$#$#$#$#$#$#$#$#$#$#$#$#$#$#$#$#$#$#$#$#$#$#$
%#$#$#$#$#$#$#$#$#$#$#$#$#$#$#$#$#$#$#$#$#$#$#$#$#$#$#$#$#$#$#$#$#$#$#$#$#$#$#$#$#$#$#$#$#$#$#$#$#$#$#$#$#$#$#$#$#$#$#$#$#$#$#$#$#$#$#$
%#$#$#$#$#$#$#$#$#$#$#$#$#$#$#$#$#$#$#$#$#$#$#$#$#$#$#$#$#$#$#$#$#$#$#$#$#$#$#$#$#$#$#$#$#$#$#$#$#$#$#$#$#$#$#$#$#$#$#$#$#$#$#$#$#$#$#$
\section*{ACKNOWLEDGMENTS}

We are very grateful to Patrik Ferrari for discussions and for providing data on the distribution tails from the TASEP simulations of Ref. \cite{CFS} at an initial stage of this work. We also thank Pavel Sasorov and Gunter Sch\"{u}tz for discussions, Andreas Schadschneider for useful comments and  Herbert Spohn for encouragement. B.M. acknowledges support from the Israel Science Foundation (Grant No. 807/16) and from the United States-Israel Binational Science Foundation (BSF) (Grant No. 2012145). J.S. acknowledges support by Deutsche Forschnungsgemeinschaft (DFG) under grant SCHA 636/8-2.
Parts of this work were done during a stay of B.M. at the Institut Henri Poincare (IHP)--Centre Emile Borel during the trimester ``Stochastic Dynamics Out of Equilibrium", and during a stay of J.S. at the Racah Institute of Physics of the Hebrew University of Jerusalem.  We thank these institutes for hospitality and support.

%!#!#!#!#!#!#!#!#!#!#!#!#!#!#!#!#!#!#!#!#!#!#!#!#!#!#!#!#!#!#!#!#!#!#!#!#!#!#!#!#!#!#!#!#!#!#!#!#!#!#!#!#!#!#!#!#!#!#!#!#!#!#!#!#!#!#!#
%!#!#!#!#!#!#!#!#!#!#!#!#!#!#!#!#!#!#!#!#!#!#!#!#!#!#!#!#!#!#!#!#!#!#!#!#!#!#!#!#!#!#!#!#!#!#!#!#!#!#!#!#!#!#!#!#!#!#!#!#!#!#!#!#!#!#!#
%!#!#!#!#!#!#!#!#!#!#!#!#!#!#!#!#!#!#!#!#!#!#!#!#!#!#!#!#!#!#!#!#!#!#!#!#!#!#!#!#!#!#!#!#!#!#!#!#!#!#!#!#!#!#!#!#!#!#!#!#!#!#!#!#!#!#!#
\appendix
%!#!#!#!#!#!#!#!#!#!#!#!#!#!#!#!#!#!#!#!#!#!#!#!#!#!#!#!#!#!#!#!#!#!#!#!#!#!#!#!#!#!#!#!#!#!#!#!#!#!#!#!#!#!#!#!#!#!#!#!#!#!#!#!#!#!#!#

%&&&&&&&&&&&&&&&&&&&&&&&&&&&&&&&&&&&&&&&&&&&&&&&&&&&&&&&&&&&&&&&&&&&&&&&&&&&&&&&&&&&&&&&&&&&&&&&&&&&&&&&&&&&&&&&&&&&&&&&&&&&&&&&&&&&&&&
%&&&&&&&&&&&&&&&&&&&&&&&&&&&&&&&&&&&&&&&&&&&&&&&&&&&&&&&&&&&&&&&&&&&&&&&&&&&&&&&&&&&&&&&&&&&&&&&&&&&&&&&&&&&&&&&&&&&&&&&&&&&&&&&&&&&&&&
%&&&&&&&&&&&&&&&&&&&&&&&&&&&&&&&&&&&&&&&&&&&&&&&&&&&&&&&&&&&&&&&&&&&&&&&&&&&&&&&&&&&&&&&&&&&&&&&&&&&&&&&&&&&&&&&&&&&&&&&&&&&&&&&&&&&&&&
\section*{Appendix A. Fluctuating hydrodynamics and TASEP-KPZ duality}
\renewcommand{\theequation}{A.\arabic{equation}}
\setcounter{equation}{0}
%\label{APPENDIX:Duality_NLFH_KPZ}
%&&&&&&&&&&&&&&&&&&&&&&&&&&&&&&&&&&&&&&&&&&&&&&&&&&&&&&&&&&&&&&&&&&&&&&&&&&&&&&&&&&&&&&&&&&&&&&&&&&&&&&&&&&&&&&&&&&&&&&&&&&&&&&&&&&&&&&
%&&&&&&&&&&&&&&&&&&&&&&&&&&&&&&&&&&&&&&&&&&&&&&&&&&&&&&&&&&&&&&&&&&&&&&&&&&&&&&&&&&&&&&&&&&&&&&&&&&&&&&&&&&&&&&&&&&&&&&&&&&&&&&&&&&&&&&

Fluctuating hydrodynamics is a convenient phenomenological tool for mapping the TASEP fluctuations onto the KPZ language. The long-time evolution of the TASEP at large scales is described by the conservation law $\partial_t\varrho(x,t)+\partial_x\mathtt{j}(x,t)=0$ \cite{JS:Kipn99}, where $\varrho(x,t)$ is the coarse-grained local density field, and $\mathtt{j}(x,t)$ is the associated current.
To arrive at the hydrodynamic description we expand the local density field around its average value, $\varrho(x,t)=\rho+u(x,t)$ and replace the local current field by the stationary current as a function of the local density, $j(\varrho(x,t))$. To account for
randomness, diffusion and conservative space-time white noise are added phenomenologically, resulting in
\begin{equation}
\partial_t u(x,t)=-\partial_x j\left(\rho+u(x,t)\right) + \mathcal{D}\partial_x^2 u(x,t) + \mathcal{B} \partial_x\xi(x,t).
\label{eq:JS:conti_I}
\end{equation}
The noise magnitude $\mathcal{B}$ and the diffusion coefficient $\mathcal{D}$ are related by the fluctuation-dissipation theorem
\begin{equation}
\mathcal{B}^2= 2\kappa \mathcal{D},
\label{eq:JS:FLucDissTheo}
\end{equation}
where
\begin{equation}
\kappa=\intop \left< u(0,t)u(x,t)\right> \text{d}x
\end{equation}
contains information about the system's space correlations and is a nonequilibrium analogue of the thermodynamic compressibility.

\begin{figure}
\includegraphics[width=0.45\textwidth]{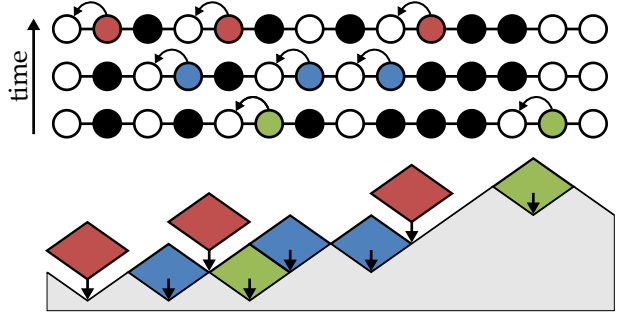}
\caption{Mapping the TASEP dynamics to a surface growth. Shown is a TASEP configuration evolving in time. The color-coded particles will hop at the next time step. Mapping a particle to an up-slope ($\CIRCLE\rightarrow\diagup$)
and a hole to a down-slope ($\Circle\rightarrow\diagdown$) one obtains, for each particle/hole
configuration, a height profile. If a particle hops to the left a diamond is added to the surface between the initial and final position of the particle.}
\label{fig:JS:TASEP_growth_map}
\end{figure}

To capture the universal behavior correctly, it suffices to expand the current-density relation up to second order \cite{JS:Spoh14}, whereas possible logarithmic corrections, arising from higher orders, are neglected \cite{JS:BT_vanB12,JS:Delf07}. One thus arrives at the equation
\begin{equation}
\partial_t u(x,t)=\partial_x\left( -j^\prime(\rho) u(x,t)-\frac{1}{2}j^{\prime\prime}(\rho)(u(x,t))^2 + \mathcal{D}\partial_x u(x,t)+\mathcal{B} \xi(x,t)\right)
.
\label{eq:JS:conti_II}
\end{equation}
Finally, performing a Galilean transformation $x\rightarrow x-j^\prime(\rho)t$ to get rid of the drift term, and denoting $\partial_xh(x,t)=u(x,t)$, Eq.~(\ref{eq:JS:conti_II}) can be transformed into the KPZ equation~(\ref{eq:KPZoriginal})
with
\begin{equation}
\nu=\mathcal{D},~~\lambda=-
j^{\prime\prime}(\rho)~~\text{and}~~\sqrt{D}=\mathcal{B}.
\label{eq:JS:KPZ_NLFH_identities}
\end{equation}
Note that the substitution $\partial_xh(x,t)=u(x,t)$ is motivated by the exact mapping of the TASEP
to a surface growth process \cite{JS:BARABASI_GROWTH,HHZ,JS:KRUG_GROWTH,JS:KRUG_Book_GROWTH,JS:TASEP_GROWTH}
that is known as the single-step
model, illustrated in Fig.~\ref{fig:JS:TASEP_growth_map}.

%##################################################################################%
%------------------------------------- FIG. 8 -------------------------------------%
%##################################################################################%
%##################################################################################%

To capture nonlinear interactions ($\lambda\not=0$) in the time-integrated current one has to account for the drift term by integrating along the directed path from $(x,0)$ to $(x+j^\prime(\rho)t,t)$  \cite{JS:FerrariFontes94,JS:MendlSpohn}:
\begin{equation}
J(x,t)=\intop_0^t  \mathtt{j}(x,s)\text{d}s -\intop_0^{j^\prime(\rho)t}  \varrho(x+x^\prime,0)\text{d}x^\prime - (j(\rho)-\rho j^\prime(\rho)) t.
\label{eq:JS:IntCurrPath}
\end{equation}
At long times the current distributions, when starting from an initial state with a given $\sigma$,  are expected to show the scaling behavior
\begin{equation}
\mathcal{P}_{t}\left(-J ,\sigma\right)  \simeq
\left(4\left|j^{\prime\prime}(\rho)\right|\kappa^{2} t\right)^{-1/3}
F^{\left(\sigma\right)}\left[\frac{J}{\left(4\left|j^{\prime\prime}(\rho)\right|\kappa^{2} t\right)^{1/3}}\right] .
\label{eq:JS:Scaling_relation_Current_Fluc}
\end{equation}
By virtue of our analytic results, the positive tail of $F^{(\sigma)}$ has the following form:
\begin{equation}
-\ln F^{(\sigma)}(x\to \infty)\simeq f(\sigma) \,x^{3/2}
\end{equation}
with $f(\sigma)$ given by Eq.~(\ref{eq:fsigma}). Expressing the interface height via the time-integrated current,
\begin{equation}
H(x,t)=-J(x,t),
\label{eq:JS:Height_Current_Identity}
\end{equation}
and using the fluctuation-dissipation theorem~(\ref{eq:JS:FLucDissTheo}) and identities~(\ref{eq:JS:KPZ_NLFH_identities})  to relate the nonuniversial scaling factors,
\begin{equation}
\frac{|\lambda|D^2}{\nu^2}=4|j^{\prime\prime}|\kappa^2
\label{eq:JS_current_scaling_factor}
\end{equation}
one can finally transform the current distribution~(\ref{eq:JS:Scaling_relation_Current_Fluc}) into the height distribution~(\ref{eq:JS:Scaling_relation_Height_Fluc}).

%&&&&&&&&&&&&&&&&&&&&&&&&&&&&&&&&&&&&&&&&&&&&&&&&&&&&&&&&&&&&&&&&&&&&&&&&&&&&&&&&&&&&&&&&&&&&&&&&&&&&&&&&&&&&&&&&&&&&&&&&&&&&&&&&&&&&&&
%&&&&&&&&&&&&&&&&&&&&&&&&&&&&&&&&&&&&&&&&&&&&&&&&&&&&&&&&&&&&&&&&&&&&&&&&&&&&&&&&&&&&&&&&&&&&&&&&&&&&&&&&&&&&&&&&&&&&&&&&&&&&&&&&&&&&&&
%&&&&&&&&&&&&&&&&&&&&&&&&&&&&&&&&&&&&&&&&&&&&&&&&&&&&&&&&&&&&&&&&&&&&&&&&&&&&&&&&&&&&&&&&&&&&&&&&&&&&&&&&&&&&&&&&&&&&&&&&&&&&&&&&&&&&&&
\section*{Appendix B. Current fluctuations in TASEP with parallel update rule}
\renewcommand{\theequation}{B.\arabic{equation}}
\setcounter{equation}{0}
%\label{APPENDIX:TASEP_Parallel_Update}
%&&&&&&&&&&&&&&&&&&&&&&&&&&&&&&&&&&&&&&&&&&&&&&&&&&&&&&&&&&&&&&&&&&&&&&&&&&&&&&&&&&&&&&&&&&&&&&&&&&&&&&&&&&&&&&&&&&&&&&&&&&&&&&&&&&&&&&
%&&&&&&&&&&&&&&&&&&&&&&&&&&&&&&&&&&&&&&&&&&&&&&&&&&&&&&&&&&&&&&&&&&&&&&&&&&&&&&&&&&&&&&&&&&&&&&&&&&&&&&&&&&&&&&&&&&&&&&&&&&&&&&&&&&&&&&
Here we present some analytical relations for the parallel-update TASEP, which allow to establish the current distributions without any adjustable parameters.
In the parallel-update TASEP all particles attempt to hop to the left at the same time with probability $q\in(0,1)$, see Fig.~\ref{fig:JS:TASEP_growth_map}. The move is disallowed if the target site is occupied.
The exact stationary probability distribution $\bar{P}(\vec{n})$ of a configuration $\vec{n}$ is given by the two-cluster approximation \cite{JS:Yaguchi86}
\begin{equation}
\bar{P}\left(\vec{n}\right)=\prod_{k=-\infty}^{\infty}P\left(n_{k},n_{k+1}\right) ,
\label{eq:JS:initial_configuration_distribution}
\end{equation}
where $n_k$ (either $0$, or $1$) is the occupation number at site $k$.
Using the Kolomogorov consistency conditions with $P(1)=\rho$ and $P(0)=1-\rho$ \cite{JS:Schadneider_STCS_book_2011}, one obtains the relations
\begin{equation}
P(0,0)=1-\rho-P(1,0),~~P(1,1)=\rho-P(1,0),~~P(0,1)=P(1,0).
\label{JS:KolomogrovConsistency}
\end{equation}
Consequently, to derive the stationary distribution one has to solve the master equation for $P(1,0)$ which reduces to a quadratic equation and yields \cite{JS:AS_PTASEP_I,JS:AS_PTASEP_II}
\begin{equation}
P_{\mathrm{st}}\left(1,0\right)=\frac{1}{2q}\left[1-\sqrt{1-4q\rho\left(1-\rho\right)}\right].
\end{equation}
With the stationary distribution at hand, one can calculate the system's current as a function of the density, its second derivative and the compressibility as
\begin{eqnarray}
j\left(\rho\right) & =&-\frac{1}{2}\left(1-\sqrt{1-4q \rho\left(1-\rho\right)}\right),\\
j^{\prime\prime}\left(\rho\right) & =&\frac{2q\left(1-q\right)}{\left[1-4q \rho\left(1-\rho\right)\right]^{3/2}},\\
\kappa & =&\rho\left(1-\rho\right)\sqrt{1-4q \rho\left(1-\rho\right)}.
\end{eqnarray}
We set the density to $\rho=1/2$, so that the drift term vanishes, $j^\prime(1/2)=0$, whereas the current-distribution scaling factor~(\ref{eq:JS_current_scaling_factor}) is at its maximum, leading to the fastest convergence to the asymptotic regime Eq.~(\ref{eq:JS:Scaling_relation_Current_Fluc}).
Translating the initial configurations, drawn from the distribution~(\ref{eq:JS:initial_configuration_distribution}), into height profiles (see Fig.~\ref{fig:JS:TASEP_growth_map}), we can express the Brownian interface diffusion constant $\tilde{\sigma}^2$ as
\begin{equation}
\tilde{\sigma}^2=\frac{P(0,0)}{P(1,0)}.
\end{equation}
As $\sigma$ is defined as the square root of the ratio of $\tilde{\sigma}^{2}$  and the diffusion constant describing the stationary state, the initial $\sigma$-states are drawn from the distribution~(\ref{eq:JS:initial_configuration_distribution}) using
\begin{equation}
P_\sigma(1,0)=\frac{1}{2}\left(1+\sigma^2\sqrt{1-q}\right)^{-1}.
\label{JS:eq:sigma_2_cluster_prob}
\end{equation}
Finally, the time-integrated current between sites $k$ and $k-1$ is
\begin{equation}
J_k\left(t\right)  =\sum_{s=0}^{t-1}n_{k-1}^{s+1}
\left(n_{k-1}^{s}-1\right)n_{k}^{s} -j(\rho)t,
\end{equation}
where $n_k^t$ is the occupation number of site $k$ at time $t$.


\begin{thebibliography}{99}
\bibitem{shifted} $H$ is obtained by subtracting from the interface height $h(x,t)$ the systematic interface motion
which results from the noise rectification by the
KPZ nonlinearity \cite{S2016,Hairer}.
\bibitem{signlambda}
Flipping the sign of $\lambda$ is  equivalent to flipping the sign of $h$.
\bibitem{numerics}  As in the previous papers on large deviations of height in the KPZ equation \cite{MKV,JKM}, we used the Chernykh-Stepanov back-and-forth iteration algorithm \cite{Chernykh}. The numerical solutions help gain intuitition about the character of the analytical solution to be found.
\bibitem{delta} At $t$ close to 1, the $\rho$-soliton rapidly transforms into the delta-function (\ref{eq:pT}), so that the solution exhibits a narrow boundary layer in time at $t=1$. This boundary layer, and another one at $t=0$, do not contribute to the action $s_{\text{dyn}}$ in the leading order in $-H\gg 1$. This feature is common for a broad class of problems dealt with with the OFM.
\end{thebibliography}
\end{document}